\documentclass[aps,twocolumn,prl]{revtex4}
\usepackage{graphicx}
\usepackage{amsfonts}
\usepackage{amssymb}
\usepackage{amsmath}

\begin{document}

\title{Quantization of a Free Particle Interacting Linearly with a Harmonic Oscillator}

\author{Thomas Mainiero}
\affiliation{Department of Physics, California Institute of Technology, Pasadena, CA 91125, USA}

\author{Mason A. Porter}
\affiliation{Oxford Centre for Industrial and Applied Mathematics, Mathematical Institute, University of Oxford, OX1 3LB, United Kingdom}
%\affiliation{Department of Physics and Center for the Physics of Information, California Institute of Technology, Pasadena, CA 91125, USA}

\begin{abstract}
We investigate the quantization of a free particle coupled linearly to a harmonic oscillator.  This system, whose classical counterpart has clearly separated regular and chaotic regions, provides an ideal framework for studying the quantization of mixed systems.  We identify key signatures of the classically chaotic and regular portions in the quantum system by constructing Husimi distributions and investigating avoided level crossings of eigenvalues as functions of the strength and range of the interaction between the system's two components.  We show, in particular, that the Husimi structure becomes mixed and delocalized as the classical dynamics becomes more chaotic.
\end{abstract}

\maketitle

PACS: 05.45.Mt, 05.45.Ða

\vspace{.1 in}

Keywords: Quantum Chaos, Mixed Dynamics, avoided level crossings, Husimi distributions

\vspace{.1 in}

\textbf{Typical classical Hamiltonians systems are neither fully integrable nor fully chaotic but instead possess mixed dynamics, with islands of stability situated in a chaotic sea.  In this paper, we investigate the quantization of a recently-studied system with mixed dynamics \cite{bievre}.  This example consists of a free particle that moves around a ring that is divided into two regions.  At the boundaries between these regions, the particle is kicked impulsively by a harmonic oscillator (in a manner that conserves the system's total energy), but the particle and oscillator otherwise evolve freely.  Although the system is not generic, its separation into regular and chaotic components also allows more precise investigations (both classically and quantum-mechanically) than is typically possible, making this an ideal example to achieve a better understanding of the quantization of mixed systems.  By examining avoided level crossings and Husimi distributions in the quantum system, we investigate the quantum signatures of mixed dynamics, demonstrating that the Husimi structures of nearby states become mixed and delocalized as chaos becomes a more prominent feature in the classical phase space.}

\section{Introduction}

Investigations of the quantization of chaotic systems have become increasingly prevalent as physicists conduct more experiments at small scales and design an increasing number of devices that exploit the physics at such scales \cite{gutzwiller,fritz,reichl,marcus,raizen}.  Experiments on quantum chaos, conducted using microwave cavities \cite{stock1,sridhar}, atom optics \cite{atomoptics,kaplan}, and other systems, have examined phenomena that are both fundamental and diverse--ranging from the decay of quantum correlations \cite{kaplan} to localization in quantum wave functions \cite{sridhar} and chaotic scattering \cite{marcus}.  Despite this wealth of research, however, it is still not entirely clear how to understand the notion of chaos in quantum mechanics.  Quantum wave functions satisfy a linear differential equation (the Schr\"odinger equation), so sensitive dependence on initial conditions and the exponential divergence of nearby trajectories--key components for defining classical chaos--cannot be used to define quantum chaos.  Nevertheless, quantum analogs of classically chaotic systems do possess identifying features, so the quantizations of chaotic systems can be distinguished from the quantizations of integrable (regular) ones. 

Typical classical Hamiltonian systems are neither fully chaotic nor fully regular; rather, they have ``mixed" dynamics (i.e., a divided phase space), with islands of stability (``KAM islands'') situated in a chaotic sea.  Because generic mixed systems are very difficult to analyze, there have been numerous attempts to construct Hamiltonian systems with mixed dynamics that allow an exact, rigorous analysis.  Previously studied examples include billiards \cite{oval,saito,mush,barnett,shroomams},
%(relevant shapes include ovals, non-concentric circles, mushrooms, and so on), 
Fermi accelerators and bouncing-ball models \cite{fermi,lich,seba90}, and kicked rotors and tops \cite{fritz,break,iomin,haake87}.  

In the present paper, we investigate the quantization of a one-dimensional free particle interacting linearly with a one-dimensional harmonic oscillator.  This recently-studied example has mixed dynamics with well-separated integrable and chaotic regions \cite{bievre}.  The clean separation between different types of behavior helps simplify comparisons between the dynamics of the classical system and that of its quantization and makes this system a very illuminating one for studying the quantization of systems with mixed dynamics.  The investigation of classical-quantum correspondences is extremely difficult for generic mixed systems, which possess an infinite hierarchy of KAM islands and intricately mixed chaotic and integrable regions.  This makes the study of identifying features of chaos in carefully-chosen examples particularly important.  
%The study of identifying features using carefully-chosen examples enables one to better understand the manifestations of chaos in quantum systems, as....

The rest of this paper is organized as follows.  First, we briefly review the classical system studied in \cite{bievre}.
 %that consists of a one-dimensional free particle interacting linearly with a one-dimensional harmonic oscillator.  
 We then quantize this system and examine its sharp and broad avoided level crossings as the relative length of the interaction versus non-interaction region is varied.  We illustrate our observations using Husimi distributions, which also allow a comparison with the classical dynamics.  Finally, we summarize our results and present additional technical details of our investigation in two appendices.

\section{The Classical System}

\begin{figure}
\includegraphics[scale=0.5]{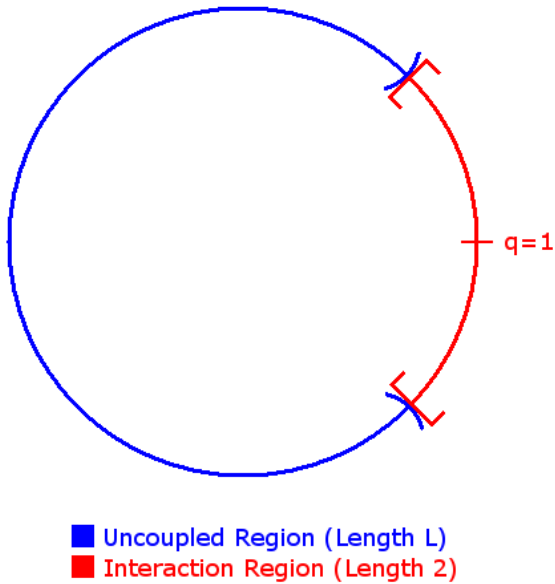}
\caption{[Color online] The configuration space of the particle.  Its position on the ring is denoted $q$.}
\label{class}
\end{figure}

Motivated by investigations of electron-phonon interactions in condensed matter physics  \cite{frolich,feynman,holstein1,holstein2,stolze91,kenkre,kenkre04,silvius,caldeira}, De Bi\`evre, Parris, and Silvius recently performed an analysis of a closed (classical) Hamiltonian system consisting of an interacting one-dimensional free-particle and a one-dimensional harmonic oscillator \cite{bievre}.  The example they investigated consists of a classical particle moving on a ring divided into two sections (see Fig.~\ref{class}): a region of length $2$ called the ``interaction region" and one of length $L$ called the ``non-interaction" (or ``uncoupled") region.  At the boundaries between these regions, the particle is kicked impulsively by a harmonic oscillator (which moves on a line rather than on the ring), but the particle and oscillator are otherwise uncoupled.  However, the harmonic oscillator oscillates about different equilibrium points in the two regions; as discussed below, the equilibrium position in the interaction region depends on the strength of the interaction.

%a classical particle of mass $m$, position $x(t)$, and momentum $p_0(t)$ moving on a ring divided into two sections (see Fig.~\ref{class}).  In one section (of length $L$), the particle is free; in the other (of length $2\sigma$), it interacts with a harmonic oscillator of mass $M$, position $X(t)$, momentum $P(t)$, and frequency $\omega$.  The system's Hamiltonian is
%\begin{equation}
	%H =\frac{{p_0}^2}{2m}+\frac{P^2}{2M}+\frac{1}{2}M\omega^2X^2-F_0X\rho(x)\,, \label{ham}
%\end{equation}
%where $F_0$ describes the strength of the interaction (which is linear in $X$).  The quantity $\rho(x)$ determines the range of the interaction.  It equals $1$ when $\left|x\right|\leq\sigma$ and $0$ when $\left|x\right|\in(\sigma,\sigma+\frac{L}{2})$, where we note that $x$ is periodic (so that $\sigma + L/2$ is identified with $-\sigma$).  In non-dimensional form, Eq.~(\ref{ham}) becomes

The (non-dimensionalized) Hamiltonian describing this system is
\begin{equation}
	H=\frac{1}{2} \left( p^2+\Pi^2+\Phi^2 \right)-\alpha\Phi\chi(q)\,,  \label{hamnod}
\end{equation}
where $p$ and $\Pi$ are, respectively, the particle and oscillator momenta, $q$ and $\Phi$ are the particle and oscillator positions, $\alpha$ describes the strength of the particle-oscillator interaction, and $\chi(q)$ takes the value $1$ in the interaction region and $0$ everwhere else.  We choose coordinates so that the interaction and non-interaction regions occur, respectively, when $q \in [0,2]$ and $q \in (2,2+L)$, where $L$ is the length of the non-interaction region and $2 + L$ is identified with $0$.  The only system parameters that can be varied are $\alpha$ and $L$.

Let's review some of the basic qualitative dynamics of (\ref{hamnod}) \cite{bievre}.  The system achieves its ground state when the particle is in the interaction region and the oscillator and particle are both at rest.  When transitioning between the interaction and non-interaction regions, the particle's momentum changes discontinuously and its position changes continuously.  When the particle enters the interaction region, the harmonic oscillator experiences an interaction force that shifts its equilibrium position 
%(about which it oscillates) 
from $0$ to $\alpha$.  (The oscillator's momentum and amplitude depend continuously on time.)  One uses conservation of energy to compute the impulse that the oscillator imparts to the particle at the transition points.    See Ref.~\cite{bievre} for further details.

%The equations of motion resulting from (\ref{hamnod}) are
%\begin{align}
%	\dot{q} &= p\,, \qquad \dot{p}=\alpha\Phi[\delta(q+1)-\delta(q-1)]\,, \nonumber  \\
%	\dot{\Phi} &= \Pi\,, \qquad \dot{\Pi}= -{\Phi}+\alpha\chi(q)\,, \label{motion}
%\end{align}
%where dots denote time derivatives. Equation (\ref{motion}) indicates that the particle behaves freely in both the interaction and non-interaction regions.  The interesting dynamics of Eq.~(\ref{motion}) arise because of what happens when a particle arrives at boundaries between the two regions (i.e., when $q=0$ and $q = 2$; see Fig.~\ref{class}).  At these locations, the particle reaches a potential barrier of height $\pm \alpha\Phi$, where the sign depends on whether the particle is coming from the interaction region $(+)$ or the non-interaction region $(-)$.  The particle bounces off the boundary elastically if its kinetic energy $p^2$ is less than the barrier height.  Otherwise, it overcomes the barrier and enters the new region with a kinetic energy of $p^2 \mp \alpha\Phi$.  Because the barrier height depends on the oscillator coordinate $\Phi$, the dynamics can be rather complicated.

%{map: I commented out the equations of motion; we don't actually need to keep that, right?}

The uncoupled Hamiltonian, given by equation (\ref{hamnod}) with $\alpha=0$, is integrable.  It possesses two independent integrals that correspond, respectively, to the energy of the harmonic oscillator and the momentum of the particle.  The latter integral leads to an $SO(2)$ rotational symmetry in the particle's configuration space.  Indeed, (\ref{hamnod}) is invariant under symplectic transformations 
\begin{equation}
	T_{A}:(p,q,\Pi,\Phi) \mapsto (\det(A)p,Aq,\Pi,\Phi),\,\,A \in O(2)\,,
\label{sym}
\end{equation}
forming a symmetry group isomorphic to $O(2)$.  However, for $\alpha \neq 0$ (i.e., the generic case), the Hamiltonian (\ref{hamnod}) no longer possesses the two integrals and is only invariant under the subset of the symplectic transformations (\ref{sym}) with either $A=1$ or $A$ a reflection about the line passing through $q=1$ and $q=1+L/2$ (see Fig.~\ref{class}).  Such transformations form a subgroup of $O(2)$ isomorphic to $\mathbb{Z}_2=\{1,-1\}$, reducing the   
%Hence, the coupling term reduces the $O(2)$ 
symmetry of the system to a parity symmetry about an axis of the ring in Fig.~\ref{class}.

A particularly interesting facet of this system is the clean separation of the integrable and chaotic regions in its phase space. Phase portraits of the system possess two characteristic integrable regions (among other structures).  The first, which exists for system energies varying from the ground-state energy to a critical positive energy, arises when the particle never leaves the interaction area.  For small positive energies outside this integrable region, the motion appears to be fully chaotic without any additional KAM structures near the boundary between the two regions \cite{bievre}.  The second integrable region is an elliptic KAM island centered on the equilibrium point that arises from the orbit in which the particle traverses each section of the ring exactly once per period.  More generally, the distinction between chaotic and integrable dynamics in this system is clear in a variety of situations.
% in which one or both regions exist.  
%This helps simplify comparisons between the dynamics of the classical system and its quantization and makes the present example a very illuminating one for studying the quantization of systems with mixed dynamics.  The investigation of such correspondences is extremely difficult for generic mixed systems, which possess an infinite hierarchy of KAM islands and intricately mixed chaotic and integrable regions.  

\section{The Quantum System}

We quantize Eq.~(\ref{hamnod}) using canonical quantization \cite{cohen-tannoudji}.  In so doing, we assume that the particle and harmonic oscillator act as bosons with no internal degrees-of-freedom (DOF) and impose the following commutation relations:
\begin{align} \label{com}
	[p,q] &= -i\,, \quad [p,\Pi] = 0\,, \quad  [\Pi,\Phi] = -i\,, \notag \\	
	[q,\Phi] &= 0\,, \quad [p,\phi] = 0\,, 
	\quad [\Pi,q] = 0\,.
\end{align}
With the coordinate-space identifications
\begin{align}
	p = -i\frac{\partial}{\partial q}\,, \quad \Pi=-i\frac{\partial}{\partial \Phi}\,,
\label{op}
\end{align}
we obtain the quantum Hamiltonian
\begin{equation}
	H=\frac{1}{2} \left( -\frac{\partial^2}{\partial q^2}-\frac{\partial^2}{\partial \Phi^2}+\Phi^2 \right) -\alpha\Phi\chi(q) \,.
\label{hamquant}
\end{equation}

\begin{figure}
%\centerline{
\includegraphics[scale=0.6]{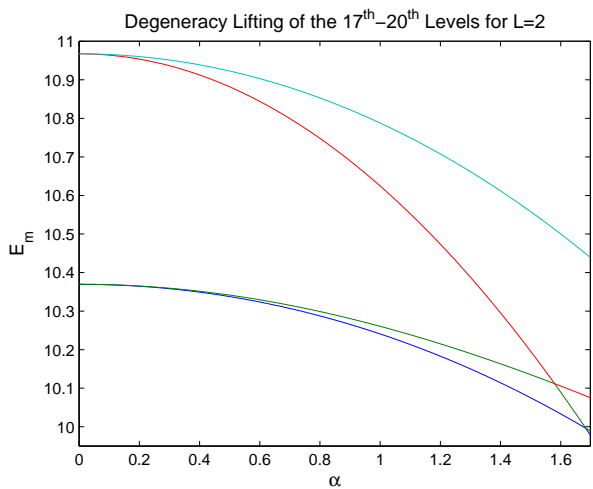} 
%}
\caption{[Color online] Lifting of the double degeneracy in the eigenenergies for $\alpha > 0$.}  
\label{split}
\end{figure}

\begin{figure}
%\centerline{
(a)\includegraphics[scale=0.5]{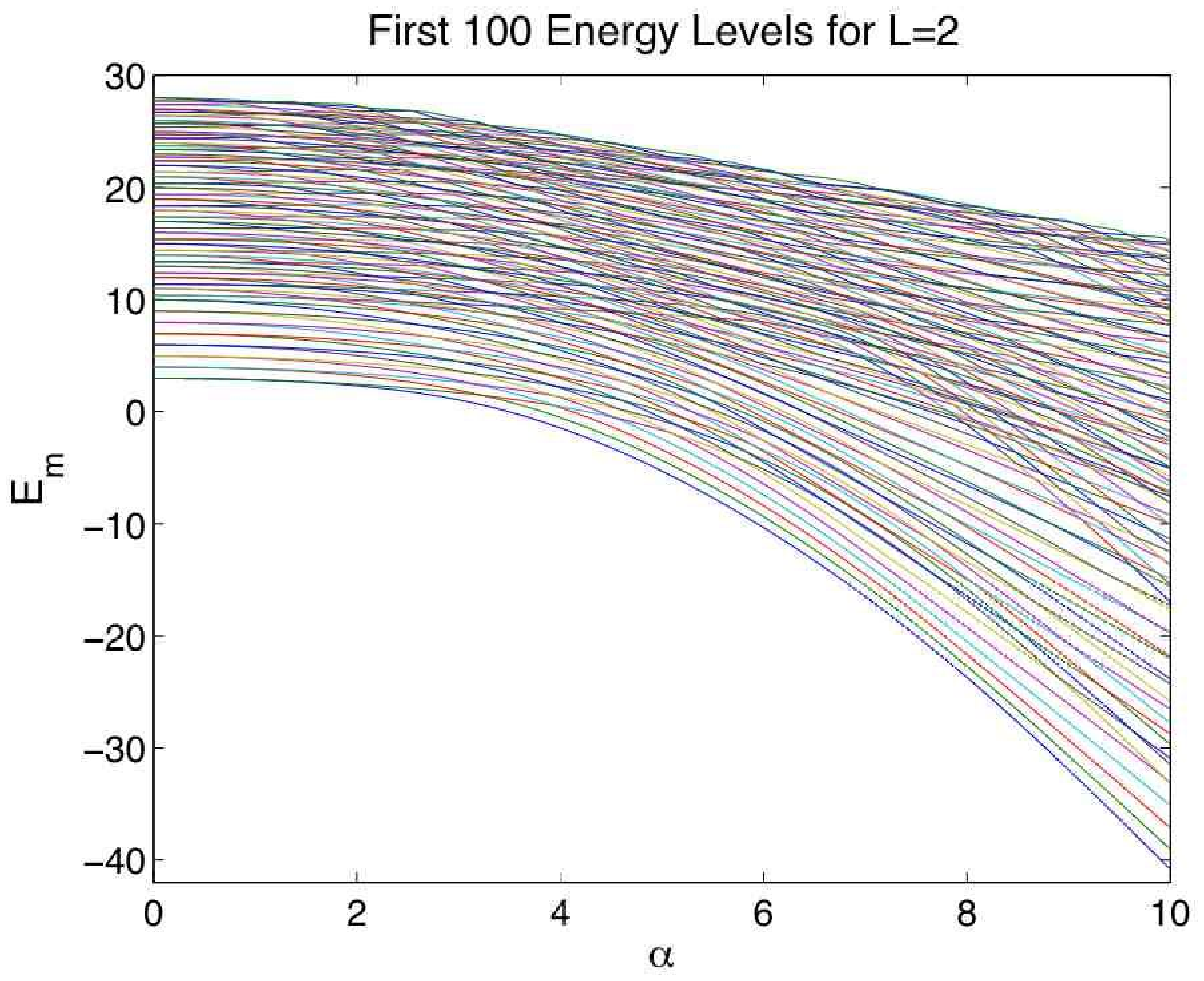}
%}
%\centerline{ 
(b)\includegraphics[scale=0.5]{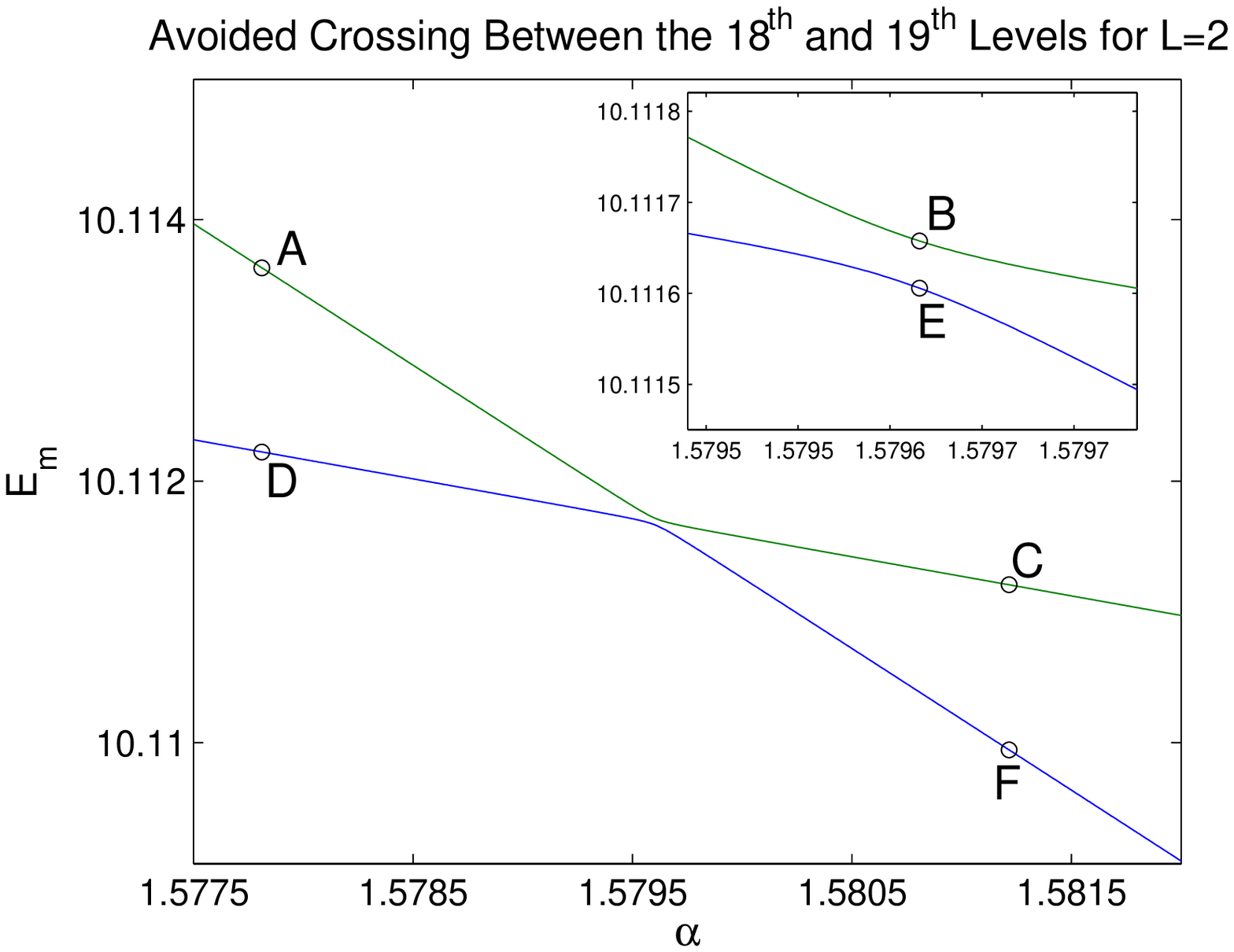}
%}
\caption{[Color online] (a) The first 100 energy levels as a function of the interaction strength $\alpha$ for Eq.~({\protect \ref{hamquant}}) with a non-interaction region of length $L=2$.  (b) Magnification of the avoided crossing between the $18$th and $19$th levels from panel (a).  The inset shows a further magnification, and the labels designate where we calculated Husimi distributions (see Fig.~\ref{h_ac}). 
}
\label{levels1}
\end{figure}

\begin{figure}
%\centerline{
(a)\includegraphics[scale=0.5]{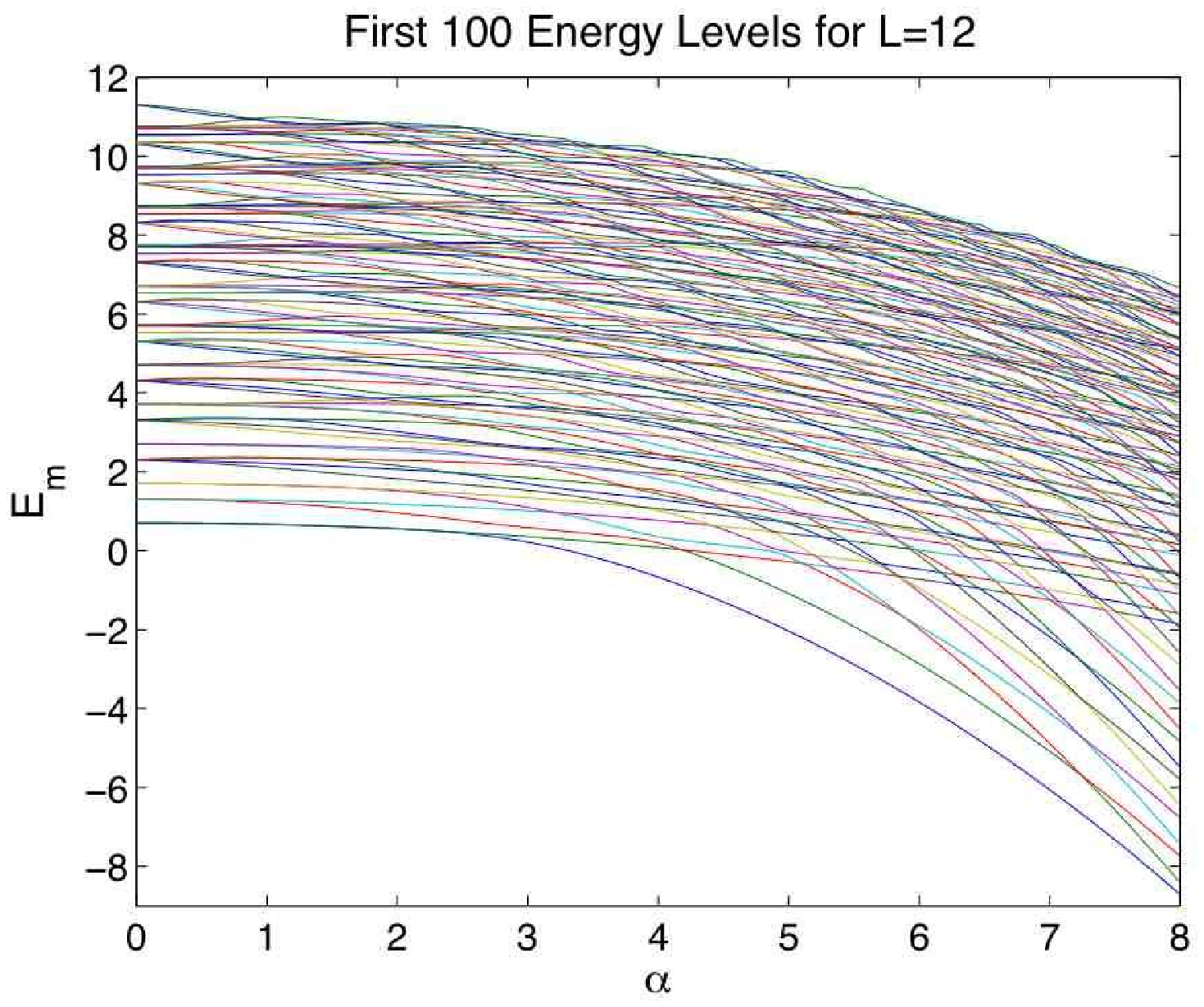}
%}
%\centerline{ 
(b)\includegraphics[scale=0.5]{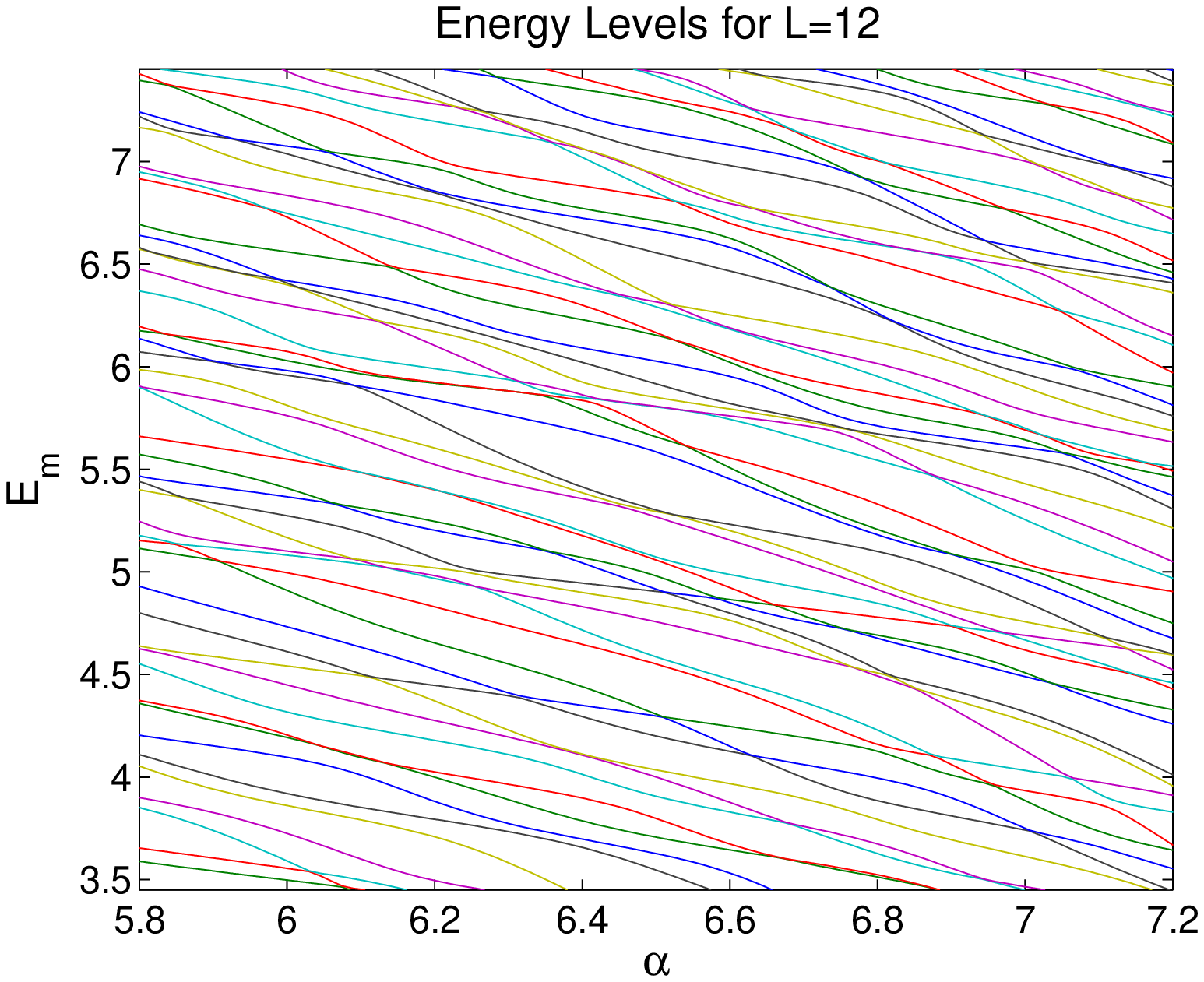}
%}
\caption{[Color online] (a) The first 100 energy levels as a function of $\alpha$ for Eq.~(\ref{hamquant}) with a non-interaction region of length $L=12$.  (b) Magnification of panel (a) illustrating broad avoided crossings.}
\label{levels2}
\end{figure}

%{map: Tom, please double check again (quadruple check?) to make sure that all the descriptions describe the current numerics and that there are no more remnants of the old numerics; also, I rearranged the figures a bit, so please also check to make sure I revised the references to these figure panels appropriately}

For the uncoupled system [Eq.~(\ref{hamquant}) with $\alpha=0$], the time-independent Schr\"{o}dinger equation $H|\psi\rangle=E|\psi\rangle$ is separable, so one just needs to determine the eigenstates of the one-dimensional harmonic oscillator (moving on a line) and the free particle confined to a ring of length $2+L$ as separate problems (both of which admit closed-form solutions).  Let $\{ |\psi^{part}_{n}\rangle \}_n$ and $\{ |\psi^{osc}_k\rangle \}_k$ denote eigenstates of the particle and the harmonic oscillator, respectively, so that $\{ |\psi^{part}_{n}\rangle \otimes |\psi^{osc}_k\rangle \}_{n,k}$ are eigenstates for the uncoupled system.
 %\footnote{Formally, $|\psi^{part}_{n}\rangle \otimes |\psi^{osc}_k\rangle$ is an element of the tensor product space ${\cal H}_{part} \otimes {\cal H}_{osc}$, where ${\cal H}_{part}$ and ${\cal H}_{osc}$ are the particle and oscillator Hilbert spaces, respectively.  One can also think of $|\psi^{part}_{n}\rangle \otimes |\psi^{osc}_k\rangle$ in terms of the coordinate-space wavefunction $\psi^{part}_{n}(q) \psi^{osc}_k(\Phi)=\langle q,\Phi | \{ |\psi^{part}_{n}\rangle \otimes |\psi^{osc}_k\rangle \}\rangle$ \cite{cohen-tannoudji}.}.  
 In fact, these states form a basis for the Hilbert space of either the coupled or uncoupled system.  We represent the Hamiltonian (\ref{hamquant}) as an infinite matrix using this basis (see Appendix I) and approximate its eigenvalues and eigenstates using those of a truncation of the matrix.  Because the eigenenergies of the particle are doubly degenerate, the eigenenergies of the full uncoupled system are also doubly degenerate.  Second-order perturbations in $\alpha$ lift this degeneracy for $\alpha>0$ and $L \neq 2$; numerical calculations indicate that this degeneracy is also lifted for $L=2$ (see Fig.~\ref{split}).

The uncoupled quantum Hamiltonian
%[(\ref{hamquant}) with $\alpha=0$] 
commutes with both the particle momentum $p$ and the Hamiltonian $H_{osc}=(\Pi^2+\Phi^2)/2$ describing an isolated harmonic oscillator; these are quantized versions of the two independent integrals of the classical Hamiltonian.  The uncoupled quantum Hamiltonian is thus integrable \cite{zhang1,zhang2}.  The symmetry-breaking of the classical Hamiltonian is mirrored in the quantum system, as $[H,p]=\alpha \Phi [\chi(q),p] \neq 0$ and $[H,H_{osc}]=\alpha \chi(q)[\Pi^2,\Phi]/2 \neq 0$ for $\alpha > 0$ and generic values of $L$.  The parity symmetry in $q$ and the time-reversal symmetry, described for the coupled classical Hamiltonian, are still present in (\ref{hamquant}), but there are no obvious continuous symmetries for $\alpha > 0$.

\begin{figure}[t]
%\centerline{
\includegraphics[scale=0.5]{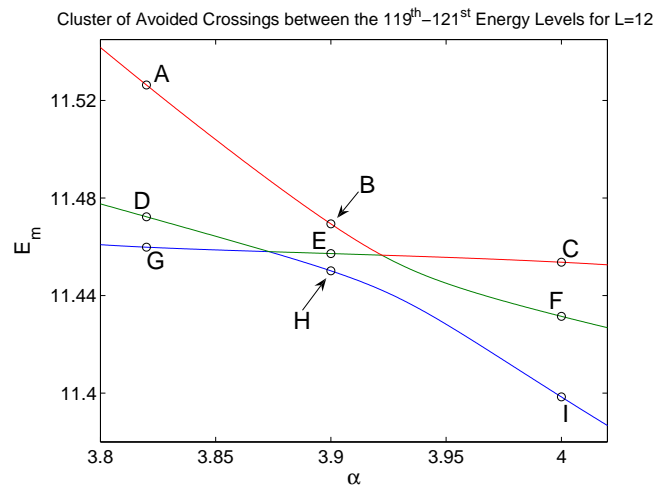}
%}
\caption{[Color online] An isolated cluster of two sharp avoided crossings and a broad avoided crossing between the $119$th, $120$th, and $121$st levels for $L = 12$. The incoming slope of the $119$th level is imparted to the outgoing slope of the $121$st.  The labels designate locations at which we calculated Husimi distributions (see Fig.~{\protect \ref{h_ac2}}).
}
\label{levels3}
\end{figure}

\section{Avoided Crossings}

As the coupling parameter $\alpha$ is varied, the eigenvalues of (\ref{hamquant}) can approach each other very closely or even cross.  If the Hamiltonian is invariant under a symmetry transformation 
%\footnote{A (quantum) Hamiltonian $H$ is invariant under a symmetry transformation $S$ if $[H,S]$=0.} 
for a certain range of $\alpha$, it can be block-diagonalized by exploiting this symmetry.  One does this by choosing each block to be invariant under the symmetry transformation.  
%By Noether's theorem, a continuous symmetry implies the existence of a conserved quantity \cite{arnold}.
%%%%Noether's theorem sometimes hints at the existence of continuous symmetries corresponding to conserved quantities \cite{arnold}.  
Energy levels belonging to different blocks can cross as $\alpha$ is varied \cite{fritz,reichl,neumann}.  On the other hand, if a quantum Hamiltonian has no symmetries other than time reversal, then such a level crossing is called an \textquotedblleft accidental degeneracy" and requires the confluence of two parameters \cite{reichl}.  In this case, most levels that approach each other end up avoiding one another instead of crossing.

Classical chaotic systems have fewer constants of motion than DOF and thus have fewer symmetries than integrable systems with the same number of DOF.  One expects the quantization of these two situations to exhibit signatures of this difference \cite{zhang1,zhang2}.  Hence, the quantization of a chaotic system should possess fewer level crossings than the quantization of an integrable one, and the presence of numerous avoided crossings between energy levels provide a signature of chaotic regions in the classical system.  Indeed, avoided level crossings are typical features of quantum chaotic systems \cite{timberlakepaper,ferez,ketz,microcav,kay}.

Figure \ref{levels1}a shows the first 100 energy levels as a function of $\alpha$ in a system with $L=2$ using an $\alpha$-step size of $1.5 \times 10^{-3}$ and a $2916 \times 2916$ truncated Hamiltonian matrix.  (The $\alpha$-step size is the distance between successive values of $\alpha$ for which we calculate eigenvalues and eigenvectors.)  One observes a multitude of
%patterns of 
apparent level crossings.
%% seem to occur in Fig.~\ref{acl2}a: we can connect each apparent crossing between two levels with the nearest apparent crossing above it to get a sequence of inscribed, continuous, concave curves.  Such curves may appear as an ``interference" pattern when the figure is viewed at a distance.
  Refining the numerical computation at some of these apparent crossings shows that they are actually avoided crossings at which the slopes of the energy level curves are exchanged.  Avoided crossings of this nature are known as \textquotedblleft sharp" avoided crossings \cite{reichl, timberlakepaper}.  In passing through such avoided crossings, the participating levels exchange their eigenstate structures, behaving as though they had entered a level crossing \cite{neumann}.  As we discuss in more detail later, we have verified numerically that this indeed occurs for our system.  Similar phenomena have also been observed in other systems, such as a sinusoidally driven particle in a square potential well \cite{timberlakepaper} and a hydrogen atom in a strong magnetic field \cite{ferez}.

Figure \ref{levels1}b shows a magnification of an avoided crossing between the $18$th and $19$th levels from Fig.~\ref{levels1}a using a refined $\alpha$-step size of $2 \times 10^{-6}$.  In general, the maximum $\alpha$-step size at which the avoided crossings in Fig.~\ref{levels1}a can be resolved is $O(10^{-6})$.  As a result, it is time-consuming to verify numerically that all of the apparent crossings are actually very sharp avoided crossings.  Although we have only observed avoided crossings,
%Although all apparent crossings that have been investigated have been shown to be avoided crossings, 
we have not ruled out that actual level crossings might occur.  Using parity symmetry, the Hamiltonian (\ref{hamquant}) can be separated into the direct sum of two blocks.  As $\alpha$ is varied for a fixed value of $L$, energy levels from different blocks might cross.

%Another notable feature of Fig.~\ref{levels1}a is that the first sixteen energy-level curves are more widely spaced than higher energy-level curves and have a much lower frequency of avoided crossings as $\alpha$ is varied.  The first sixteen energy-level curves accordingly oscillate much less as a function of $\alpha$ than do the higher energy levels.  This is unsurprising, as in the uncoupled system with $L=2$, the first sixteenh levels of the oscillator all have a lower energy than that of the ground state of the particle.  This gives the levels of the system their wide spacing, which is preserved as $\alpha$ increases.  The wide spacing prevents close encounters between these levels, in contrast to what happens for higher-energy levels.  As $L$ increases, however, the ground-state energy of the particle drops quadratically in $L$ and one does not expect this behavior to hold.

Figure \ref{levels2}a shows the first 100 energy-level curves for $L=12$.  Our numerical computations, for which we used a $4096 \times 4096$ truncated Hamiltonian matrix, verify that the frequency of avoided crossings increases as $L$ grows.  Even more interesting are the broad avoided crossings, which become more prevalent as $\alpha$ increases.  Figure \ref{levels2}b shows energy-level curves for $\alpha \in [5.8, 7.2]$.  The $\alpha$-step size for which the sharp avoided crossings in this figure can be resolved is $O(10^{-3})$, about one thousand times larger than that required to resolve the sharp avoided crossings in Fig.~\ref{levels1}a.  We have also observed that the prevalence of broad avoided crossings increases as $L$ increases.  
%{TM: I added the previous sentence as I removed something similar in previous iterations but it is needed for the section \textit{Signatures of Chaos}.}   
In general, broad avoided crossings tend to occur in nearly isolated clusters in which only a subset of the initial slopes of the participating levels are exchanged after the sequence of crossings in the cluster \footnote{We use the term \textquotedblleft isolated cluster" to designate a sequence of nearby avoided crossings whose participating energy levels are nearly unaffected by other levels in their \textquotedblleft incoming" (as a function of $\alpha$) and outgoing regions.}.  This behavior is caused by the presence of ``overlapping" avoided crossings in the cluster (by ``overlapping," we mean that there are ranges of $\alpha$ values for which multiple avoided crossings in the cluster are taking place).  Such clusters usually induce nontrivial structural exchanges between the participating eigenstates and are suspected to be an indication of chaos \cite{marcus1,marcus2}.  Indeed, prior work on a sinusoidally driven particle in a square potential well \cite{timberlakepaper} has indicated that broad avoided crossings produce superpositions of eigenstate structure rather than complete exchanges.  We will use the term \textquotedblleft mixing" to refer to such superpositions.  

An example of an isolated cluster of avoided crossings is shown in Fig.~\ref{levels3}.  This cluster occurs for $L=12$ and shows the 119th--121st levels as they experience two sharp avoided crossings and one broad one.  A broad avoided crossing begins between the 120th and 121st levels, but as $\alpha$ is increased, two sharp avoided crossings cause structural exchanges that induce a broad avoided crossing between the 119th and 120th levels.  The figure gives the appearance of a broad avoided crossing overlapping with the second sharp avoided crossing. 
%{TM: I replaced ``occuring during'' with ``overlapping'' in the previous sentence}.  
Consequently, the slopes of the participating levels in the second sharp avoided crossing are not completely exchanged.  We have observed that such behavior often arises in avoided crossings (one of which is almost always broad) that occur simultaneously as $\alpha$ is varied.  While the complex interactions between energy levels (as a function of $\alpha$) may make it impossible to precisely identify broad versus sharp avoided crossings in isolated clusters, approximate distinctions such as that just discussed allow one to classify these interactions and determine their role in structural exchanges between eigenstates (see the discussion below).

\begin{figure}[t]
\centerline{
(a)\includegraphics[scale=0.3]{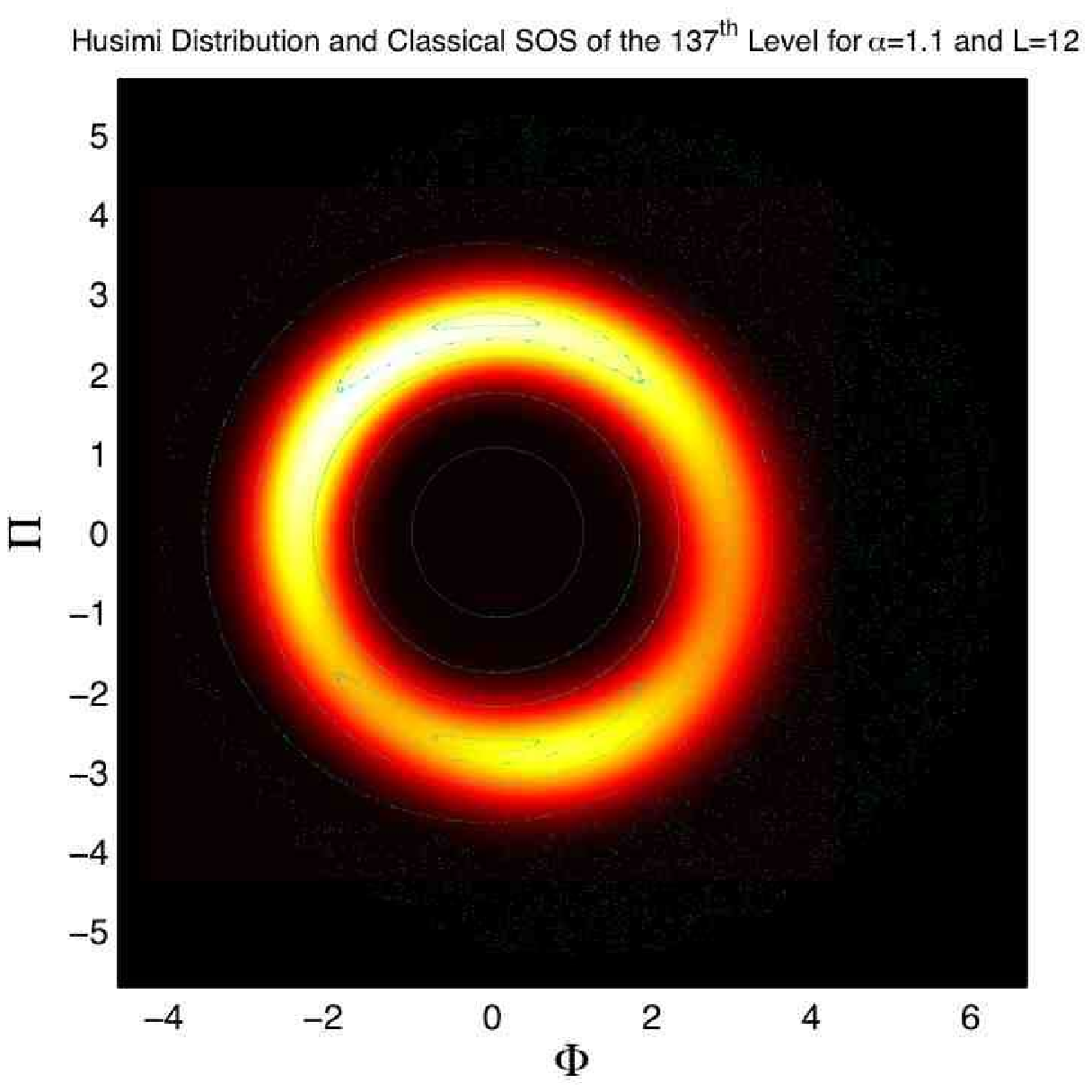}
(b)\includegraphics[scale=0.3]{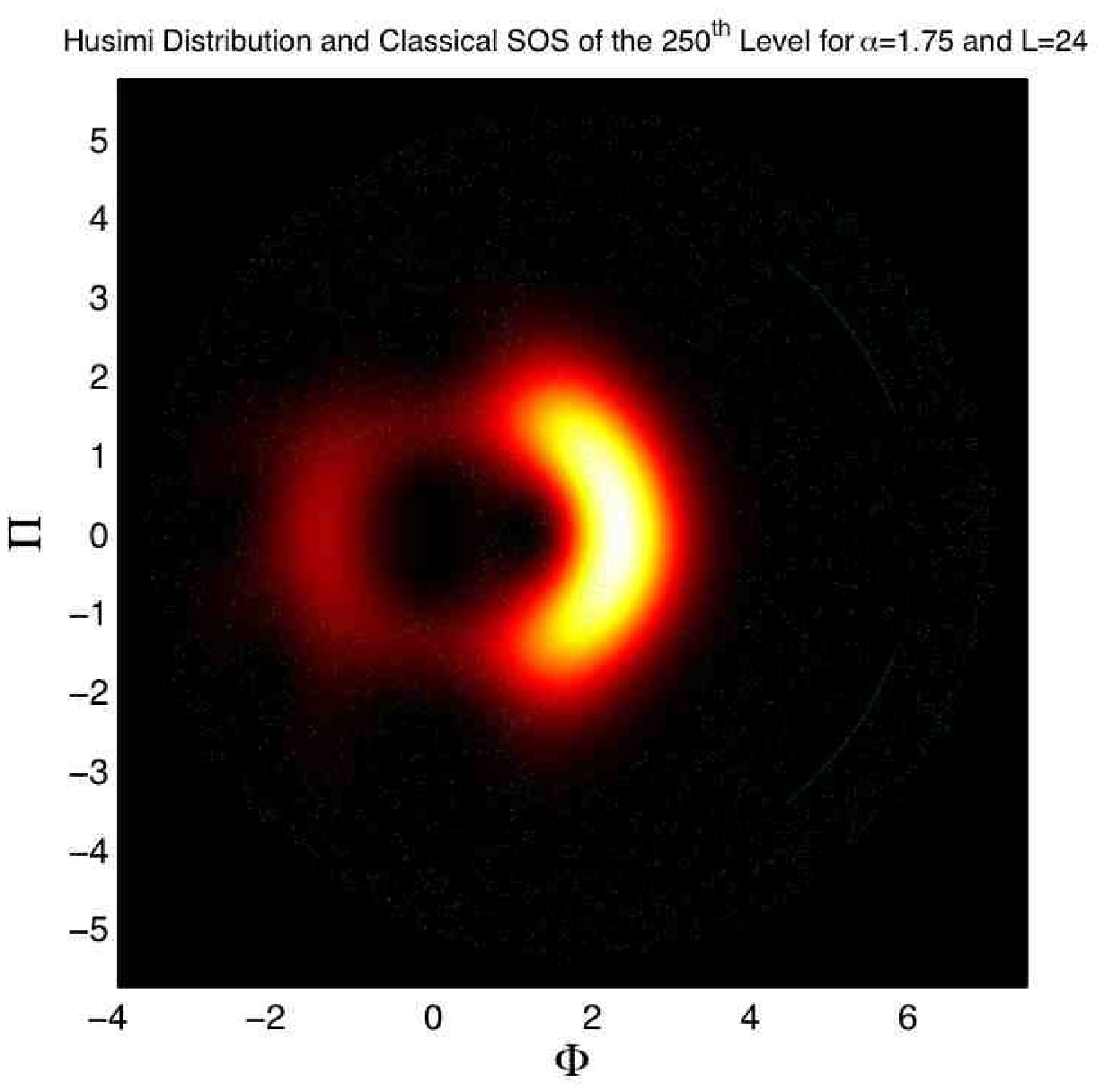}
}\caption{[Color online] (a) Husimi distribution localized around quasiperiodic orbits in a KAM island.  Lighter regions have higher probabilities and black regions are ones with zero probability.  The classical SOS is plotted in turquoise.  To facilitate the comparison between the quantum and classical dynamics, we include plots of a few of the orbits in the integrable region.  (b) Husimi distribution localized around multiple KAM islands.
}
\label{Husimi_1}
\end{figure}

\begin{figure}[t]
\centerline{
(a)\includegraphics[scale=0.3]{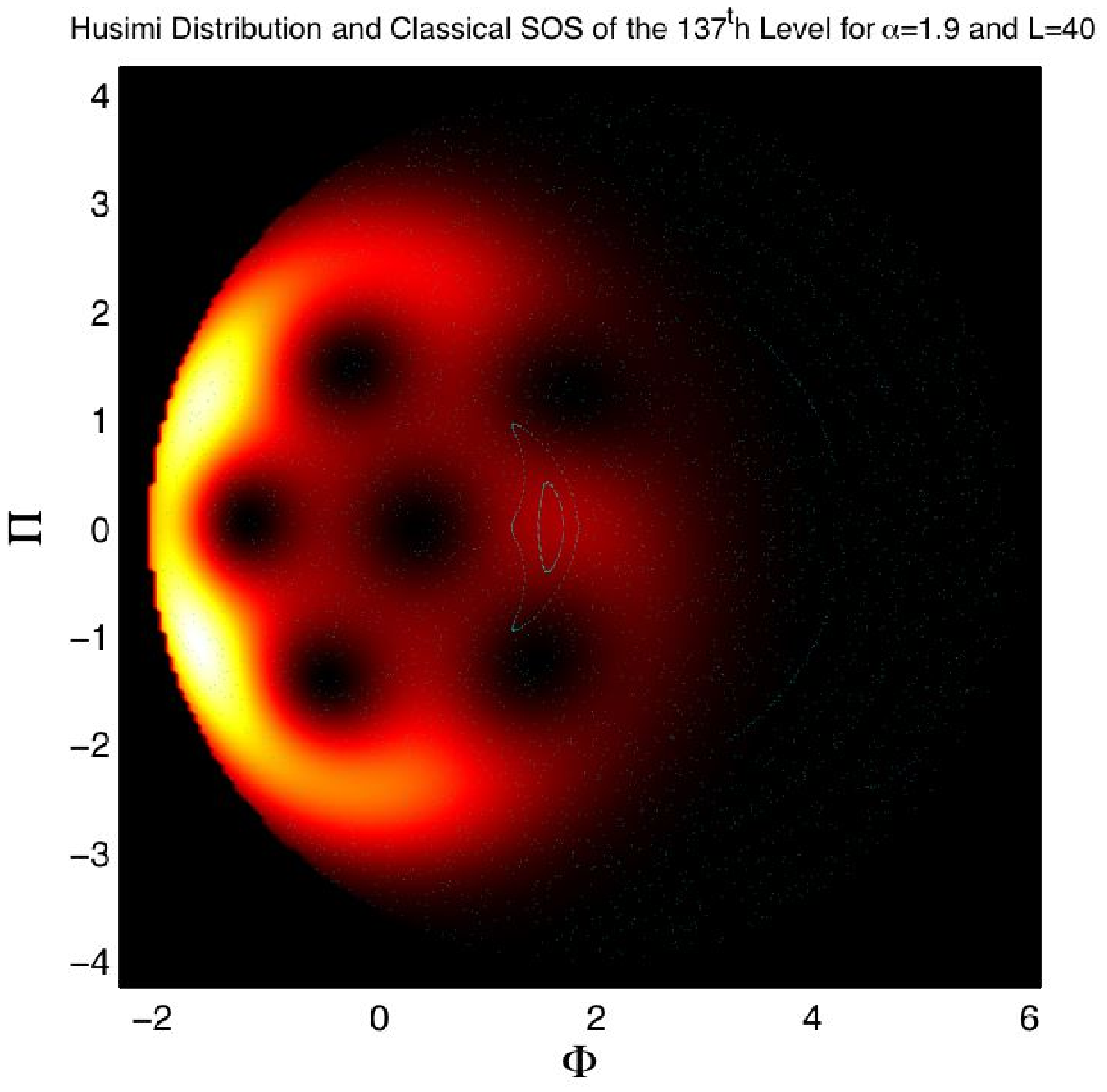}
(b)\includegraphics[scale=0.3]{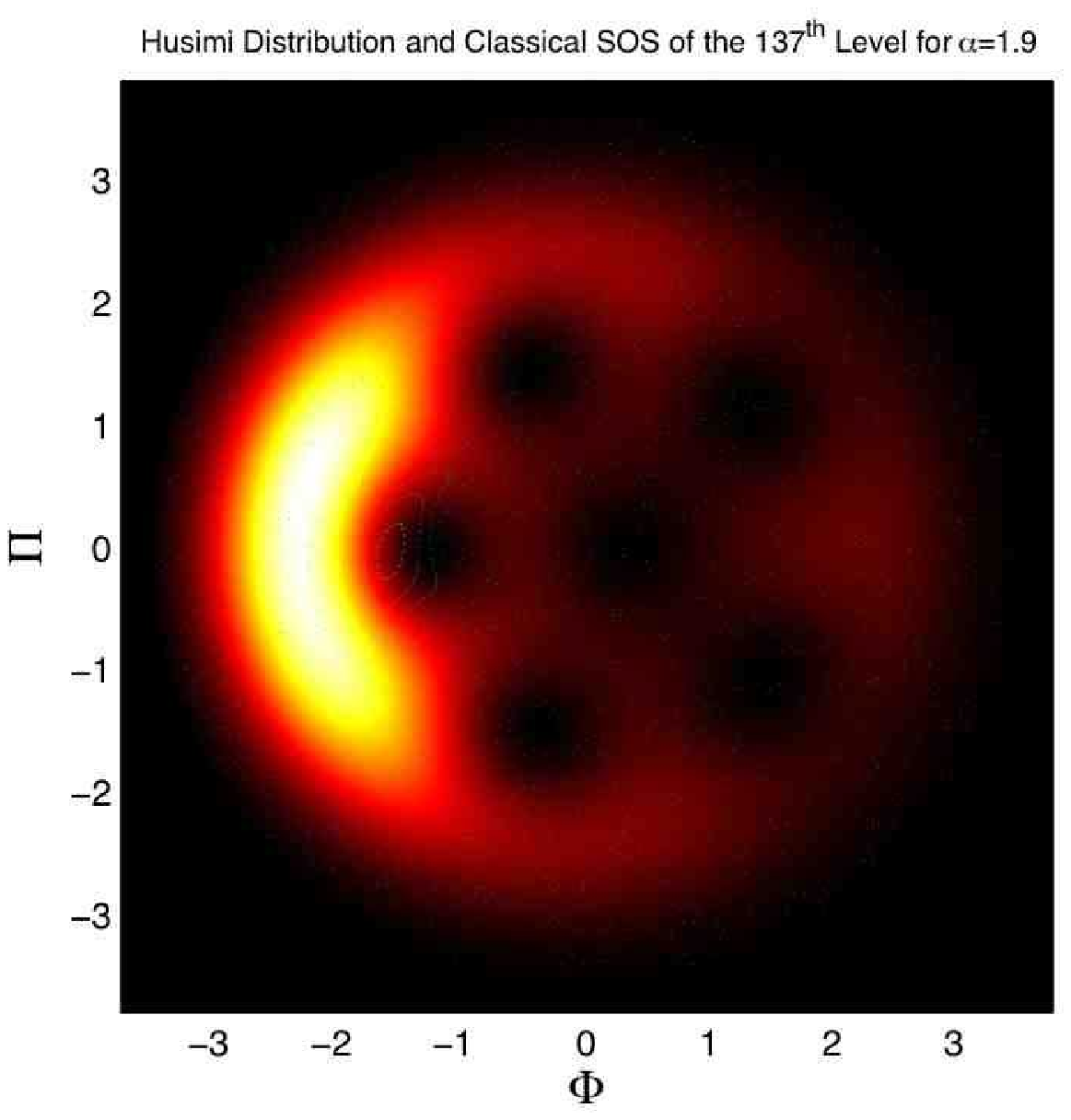}
}
\caption{[Color online] (a) Husimi distribution delocalized throughout the chaotic sea on the left half of the plot. (b) Husimi distribution and classical SOS for the chaotic eigenstate in (a) taken on the surface defined by $q=2+L/4=12$ (one fourth of the way into the non-interaction region).
}
\label{Husimi_2}
\end{figure}

One can gain insight into the energy level dynamics for highly excited states (i.e., in the deep semiclassical regime) using level spacing statistics \cite{gutzwiller,stockmann}.  Once desymmetrized, completely integrable systems obey Poisson statistics and completely chaotic ones obey Wigner statistics.  The quantization of mixed systems with well-separated integrable and chaotic regions (such as the present one) have energy level statistics that interpolate between these two extremes, with a distribution of a precise form that is conjectured to follow Berry-Robnik statistics \cite{br1,robnik93,prosen02}.  In the present paper, we focus on eigenstate structures rather than semiclassical dynamics, and we accordingly investigate Husimi distributions below and leave the analysis of energy level statistics for a future publication.

\section{Husimi Distributions}

Although there is no equivalent of classical phase space trajectories in quantum mechanics, there are suitable analogs.  In particular, the Husimi distribution is often used in the study of quantum chaos \cite{gutzwiller,weissman82}.  Given a quantum state $|\psi\rangle$, its Husimi distribution $H_{\psi}(p,q)$ is defined by the projection of $|\psi\rangle$ onto a coherent state $|\psi_{(p,q)}\rangle$ localized around $(p,q)$: $H_{\psi}(p,q)\propto|\langle\psi_{(p,q)}|\psi\rangle|^2$.  For a system with Euclidean topology, a coherent state localized at $(p,q)$ is a Gaussian with position-space representation localized around $q$ and momentum-space representation localized around $p$.  The system (\ref{hamquant}) possesses a cylindrical phase-space topology, as the particle position $q$ is a periodic variable and the momentum $p \in \mathbb{R}$.
%\footnote{In Eq.~(\ref{hamquant}), $p \in \mathbb{R}$ and $q \in (2+L) \mathbb{S}^1$, where $\mathbb{S}^1$ is the circle with unit circumference.  This gives a phase space 
%%[$(q,p)$-space] 
%topology of $(2+L) \mathbb{S}^1 \times \mathbb{R}$, which is a cylinder.}.  
%{map: if it's the unit circle, you need to divide by 2\pi in the notation you used}
In Appendix II, we construct the coherent state for this topology from the Euclidean coherent state \cite{spina}.

Coherent states provide excellent quantum analogs of classical particles when visualized as wave packets that minimize the position-momentum uncertainty product.  
%(For a one-dimensional system with position $q$ and momentum $p$, the Gaussian wave-packet minimizes the product $\Delta q \Delta p$.)  
The projection onto these particle-like states can thus be viewed intuitively as a sort of classical smearing.   One then interprets the Husimi distribution as a probability distribution in phase space \cite{backer}, allowing one to understand the dynamics of a quantum system in an analogous manner to phase portraits (in particular, surfaces of section \cite{lich,wiggins}) of its classical counterpart.  See \cite{backer,perotti,chang1,chang2} for additional discussions and applications.  For the classical system (\ref{hamnod}), we plot Poincar\'e surfaces of section (SOS), for $p > 0$ and various values of $q$ (often $q = 1$), on the projection of phase space onto the harmonic oscillator's coordinates [i.e., on $(\Phi,\Pi)$-space] \cite{bievre}.  As discussed in Appendix II, we calculate Husimi distributions for eigenstates of the Hamiltonian (\ref{hamquant}) and project them onto the quantum oscillator phase space.  Using the eigenstate energy and the same values for the system parameters $\alpha$ and $L$, we compare the Husimi distributions to the corresponding classical SOS.

\begin{figure}[t]
\centerline{
(a)\includegraphics[scale=0.3]{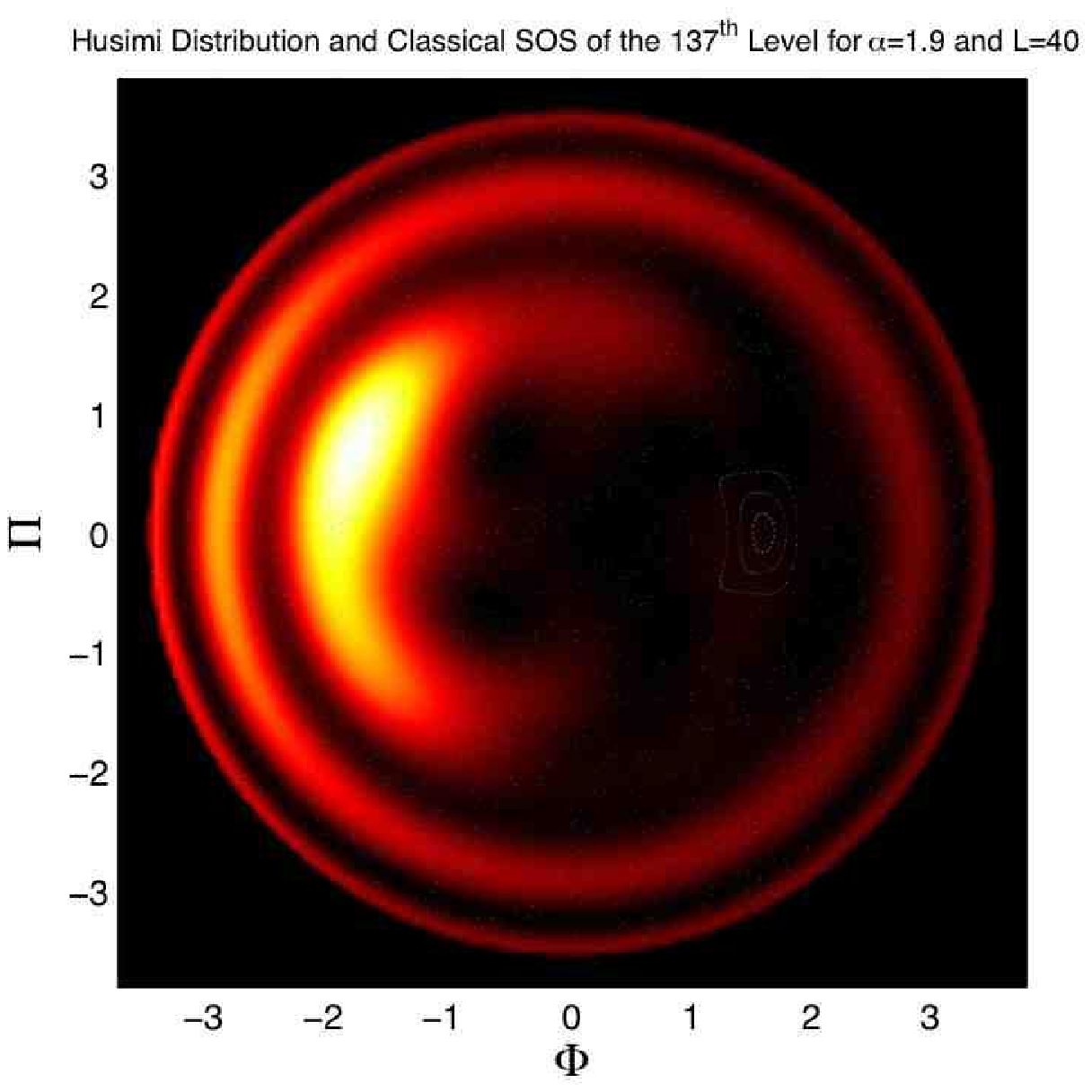}
(b)\includegraphics[scale=0.3]{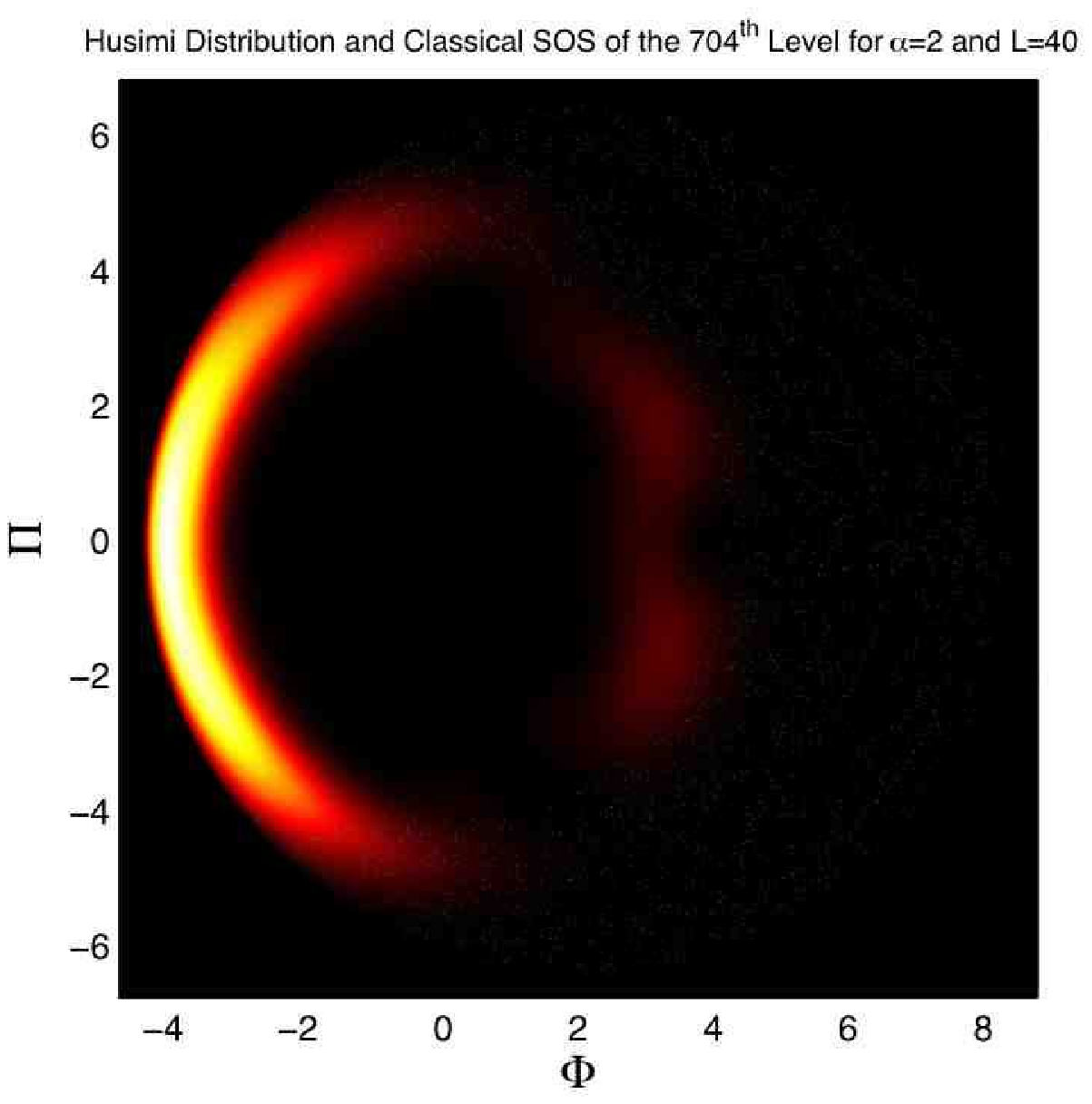}
}
\caption{[Color online] (a) Husimi distribution and classical SOS for the chaotic eigenstate in Fig.~\ref{Husimi_2}a taken on the surface defined by $q=2+L/2=22$ (in the middle of the non-interaction region).
(b) Husimi distribution showing both chaotic and integrable features while having only one connected component.  The structure is relatively delocalized throughout the enclosed area on the left side of the available phase space.  
%{map: Tom, what exactly is being depicted in panel (b); which surface of section?  this needs to be specififed further}
}
\label{Husimi_3}
\end{figure}

\subsection{Localization Around Classical Features}

To investigate the signatures of chaotic and integrable dynamics on the quantum system, we compare quantum Husimi structures to corresponding classical SOS \cite{perotti,gutzwiller,backer,timberlakepaper}.  As expected, eigenstates sometimes show localization around distinctive features of an SOS.  A prominent manifestation that often occurs is a strong localization around KAM islands, which is sometimes accompanied with Husimi density throughout the chaotic sea.  
%We found numerous examples of localization around KAM islands.  
Figure~\ref{Husimi_1} shows two such states: In panel (a), the Husimi structure is strongly localized around quasiperiodic orbits in an integrable region of the classical SOS; in panel (b), the structure is strongly localized in and around multiple KAM islands.  In this figure, we display the Husimi distribution as a contour plot that varies from black (zero probability) to white (about $10^{-3}$), with lighter regions having higher probability than darker regions.  We overlay the classical SOS in turquoise.  We observe that the quantum manifestation of a classical integrable region is a Husimi structure strongly localized inside the region.  By analogy with the classical dynamics, we refer to such states as ``regular."

%We also observed states that were not as clearly localized in integrable regions but were instead localized in nearby regions possessing the same characteristic shape (see, for example, Fig.~\ref{Husimi_1}b).

Figure \ref{Husimi_2}a shows an eigenstate whose Husimi structure is located in the chaotic sea of its corresponding classical SOS (particularly around the edge on the left half of the SOS).  We observe delocalization of the distribution throughout the chaotic sea, providing a signature of the predominantly chaotic dynamics of the classical SOS.  By analogy with the classical dynamics, we refer to quantum states with such structures as ``chaotic."  For Husimi distributions taken at $q=1$, we have only observed structures localized around the left half of the available phase space.  However, distributions constructed for $q$ values in the non-interaction region of phase space have structures with positive density throughout the available phase space.  Figures~\ref{Husimi_2}b and \ref{Husimi_3}a, for example, show Husimi distributions (and their corresponding classical SOS) for the chaotic state in Fig.~\ref{Husimi_2}a taken on the surfaces defined by $q=2+L/4=12$ (a fourth of the way into the non-interaction region) and $q=2+L/2=22$ (the middle of the non-interaction region), respectively.  These Husimi distributions still display a strong localization on the left but nevertheless have a significant density throughout the available phase space.

The enhanced density around the left edge of the chaotic sea is present for all Husimi distributions with chaotic structures.  We observe this as well for Husimi distributions taken in the non-interaction region.
 %as shown by Figs.~\ref{Husimi_2}b and \ref{Husimi_3}a.  
 As illustrated in Fig.~\ref{Husimi_3}b, this occurs even for large energies, for which the chaotic region in the left half of the SOS retreats to the edge of the available phase space.  %{map: you write that this occurs for fig 7a, but that's in the interaction region; which panel do you mean?  fig 8a?}
 The location of this enhanced density coincides with the left boundary of the intersection between the available oscillator phase space in the interaction region and the available oscillator phase space in the non-interaction region.  As discussed in detail in Ref.~\cite{bievre}, the portion of oscillator phase space corresponding to the interaction region is given by a disk-shaped region of radius $\sqrt{2E+\alpha^2}$ centered at $(\alpha,0)$ and that corresponding to the non-interaction region is given by a disk-shaped region of radius $\sqrt{2E}$ centered at the origin.  Comparing Figs.~\ref{Husimi_2}a and \ref{Husimi_2}b, we observe that this intersection in classical phase space occurs for $\Phi \approx -2$ between $\Pi \approx  -2$ and $\Pi \approx 2$.  In both plots, the Husimi density is clearly enhanced in this area, in correspondence with the dynamics of the classical system.  
 % It seems, therefore, that this density enhancement arises from the corresponding classical dynamics.

%{TM: Is this description of enhancement around the intersection region more clear?}

%This may be due to continuity of the Husimi distribution as the particle position $\bar{q}$ (see Appendix II) is varied from the the non-interaction region of phase space to the interaction region in which our classical SOS are taken.  

%The (disk-shaped) non-interaction region in phase space partially intersects the left half of the classical SOS for the majority of eigenenergies in the parameter ranges we studied (we varied the uncoupled length $L$ from 2 to 40). \cite{bievre}.
%This KAM island lies on the left hand side of the classical SOS.  
%When $L$ is larger, the system's ground state has a lower energy and a higher density of states.  More complex features (i.e., hierarchies KAM islands throughout the phase space) sometimes exist for these lower energies and one might concomitantly observe a more complicated Husimi structure.

In Fig.~\ref{Husimi_3}b, we show a Husimi distribution that contains a mixture of regular and chaotic features.  Although the regular and chaotic structures appear to be disconnected at first glance, one can see upon very close inspection that they are actually connected by small ``bridges" with positive Husimi density.  %{map: Tom, where are these bridges?  I can't tell if that part is black or has positive density; I'd like to go over this briefly in person to make sure I understand this; I zoomed this to 500\% and I still wasn't sure if there were any bridges there}
Such bridges appear more prominently between the regular structures in Fig.~\ref{Husimi_1}b.   The bridges that are often observed between Husimi structures localized in different classical phase space regions (for example, between integrable and chaotic regions) serve as channels through which density continuously flows as $\alpha$ is varied through an avoided crossing.  We did not observe any Husimi distributions in which the regular and chaotic components were completely disconnected. 

In concluding this subsection, we remark that there is a considerable body of work on ``regular" versus ``chaotic" eigenfunctions.  According to Percival's conjecture, eigenfunctions localize either on integrable or chaotic regions of the underlying classical phase space, so that the partial level density of regular states
%likelihood of observing an eigenstate localized in a particular integrable region (of the eigenstateÕs corresponding classical phase portrait) 
is given approximately by the fraction of volume of the integrable region relative to the available phase space \cite{percival,backer2,barnett}.  A recent paper by Marklof and O'Keefe \cite{marklof} contains rigorous results on such an extension of quantum ergodicity theory to a general class of quantum unitary maps whose underlying classical system has a divided phase space.  
%To our knowledge, the only other rigorous results for quantum ergodicity are for purely chaotic systems, and 
It would be extremely interesting to obtain similar results for mixed systems like the present one.

%{map: Tom, what do you think of the above phrasing?  I checked it with Alex Barnett and he thinks it's good (and he's an expert, so I think we're in good shape on this now)}

\subsection{Exchange and Mixing of Husimi Structures at Avoided Crossings}

\subsubsection{Sharp Avoided Crossings}

\begin{figure}[t]
\centerline{\includegraphics[scale=0.55]{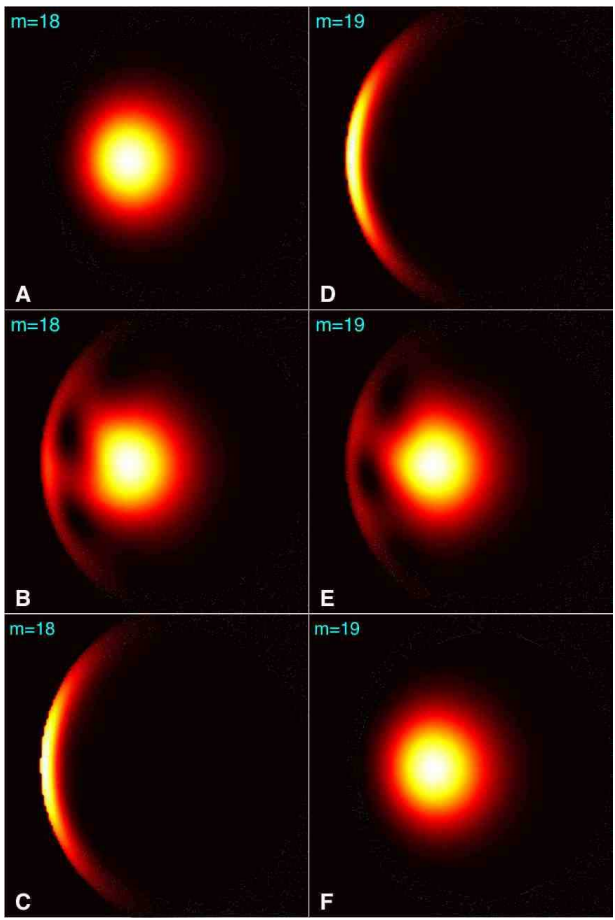}} 
\caption{[Color online] Husimi structure exchange through the sharp avoided crossing shown in Fig.~{\protect \ref{levels1}}b.  The left and right columns show the Husimi distributions of the $18$th and $19$th levels, respectively.  The harmonic oscillator momentum $\Pi$ is on the vertical axis and the oscillator position $\Phi$ is on the horizontal axis.  During the structure exchange, probability flows continuously from the integrable region to the chaotic region for the $18$th eigenstate (and vice-versa for the $19$th).}
\label{h_ac}
\end{figure}

In Fig.~\ref{h_ac}, we depict Husimi distributions for the $18$th and $19$th energy levels of (\ref{hamquant}) with $L = 2$.  This plot reveals the structural changes that occur as the two levels encounter the avoided crossing in Fig.~\ref{levels1}b.  In the top panels, one can see that the $18$th level is a regular state, whereas the $19$th level is chaotic.  The middle panels give snapshots near the closest point of the encounter.  Here, the Husimi distributions appear as mixtures of the two initial distributions.  Additionally, the regular and chaotic portions of the structure are connected.  As $\alpha$ is varied, Husimi probability flows continuously between integrable and chaotic regions.  The bottom panels, depicting snapshots from long after the encounter, show that the two levels have completely exchanged their structure through the avoided crossing, leaving the aggregate Husimi structure unchanged.  This provides an example of a smooth exchange of character in a sharp avoided crossing, which has also been observed in other quantum chaotic systems \cite{timberlakepaper,holder}.  The avoided crossings that we have observed between chaotic and regular states and between two chaotic states have all been sharp ones.

\begin{figure}[t]
\centerline{\includegraphics[scale=0.4]{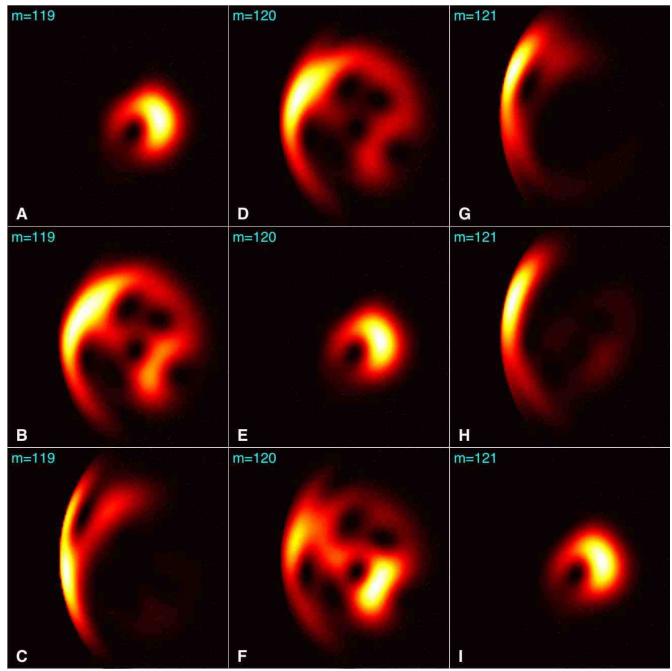}}
\caption{[Color online] Mixing of Husimi structures through the avoided-crossing cluster of Fig.~{\protect \ref{levels3}}.  The left, center, and right columns show the Husimi distributions of the $119$th, $120$th, and $121$st levels, respectively.  The axes are as in Fig.~\ref{h_ac}.
}
\label{h_ac2}
\end{figure}

\subsubsection{Broad Avoided Crossings}

As described previously, broad avoided crossings in the energy-level curves typically occur in nearly isolated clusters in which the initial slopes of the curves are not fully exchanged after the sequence of avoided crossings.  This leads to a ``mixing" of Husimi structures rather than a complete exchange \cite{timberlakepaper}, which we have verified is the case for the Hamiltonian (\ref{hamquant}). Figure~\ref{h_ac2} shows an example of such a mixing in the cluster of avoided crossings from the $119$th through the $121$st energy levels from Fig.~\ref{levels3}.  The localization of the initial Husimi structure of the $119$th level indicates that this eigenstate is regular. The $120$th level is localized near an integrable region, although there is a significant Husimi density in the chaotic sea, and the $121$st level displays a strong chaotic localization with a slight density near the integrable region. 
%{TM: I changed the previous sentence as it did not reflect the new numerics} 
The $119$th and $120$th levels leave the avoided-crossing cluster with Husimi structures that appear as mixtures between the initial Husimi structures of the $120$th and $121$st levels.  The $121$st state, however, leaves with a Husimi structure nearly identical to the initial structure of the $119$th state.  These observations are consistent with the slope exchanges in Fig.~\ref{levels3}.  This mixing causes the Husimi distributions of the $119$th and $120$th levels to delocalize after the avoided-crossing cluster.  Thus, in contrast to sharp avoided crossings, broad avoided-crossing clusters play a significant role in modifying the aggregate Husimi structure as the coupling strength $\alpha$ increases.  In particular, such clusters appear to mix the Husimi structure of individual eigenstates.

%Because broad avoided crossings become increasingly prominent (as compared to sharp avoided crossings) for larger $L$, one expects the aggregate Husimi structure to depend more strongly on the coupling strength $\alpha$ as $L$ is increased (i.e., as the non-interaction region becomes larger).

\subsubsection{Localization and $\alpha$-dependence of Avoided Crossings}

%For the non-interaction length $L=2$, Husimi distributions have a strong chaotic or regular identity, as they are almost completely localized in one of the two types of regions.  This system's avoided crossings can be resolved by $\alpha$-step sizes $O(bg 10^{-6})$.  As $L$ is increased, the $\alpha$-step size at which even the sharpest avoided crossings can be resolved tends also to increase.  

We have observed that sharp avoided crossings involve an interaction either between two chaotic levels or between a level localized primarily in an integrable region (with only slight localization in the chaotic region) and one localized primarily in a chaotic region.  However, broad avoided crossings such as those in Fig.~\ref{levels3} tend to involve interactions between levels with significant localization in both integrable and chaotic regions.

%{map: is there some sort of transition we can put right before this last sentence?  it would be nice if it doesn't appear as a non sequitar; Tom, do you have any suggestions for either a transition sentence or a place to move this sentence? }  

We have also seen that broad avoided crossings become more prevalent as $\alpha$ increases (see, for example, Fig.~\ref{levels2}a).  Because clusters of broad avoided crossings lead to mixing of the Husimi structures of the participating levels, one expects that mixing among eigenstates should also become more prominent for larger values of $\alpha$.  Additionally, the $\alpha$-step size required to resolve avoided crossings (i.e., to confirm that they are avoided crossings rather than actual level crossings) appears to be correlated with the initial localization of the Husimi distributions of the participating energy levels.
%One can investigate this phenomenon quantitatively by comparing the resolution of avoided crossings with quantitative measures of localization \cite{fritz}, thereby obtaining a quantitative distinction between sharp and broad avoided crossings and establishing additional connections to the classical dynamics.
%{map: I don't understand the relation between the resolution and the quantitative localization measures; I removed this comment, but let me know if there is something more precise/less speculative we can put here; I think it should just be removed}
Excluding avoided crossings between two chaotic states, the maximum step size that is sufficient for resolution seems to increase as the localization of the participating levels in similar regions of the classical SOS increases.  Because adjacent eigenstates with energy-level curves that are nearly parallel over some range of $\alpha$ can be interpreted as an extremely long-range avoided crossing (with a very large step size sufficient for resolution), they provide a limiting case of the above situation.  Such examples arise when the participating eigenstates are localized to almost the same extent in the same integrable region.

%{TM: I reordered this a bit and changed the wording.  Should we place the comment about the possibility of quantitative results or leave it as is?}

%{TM: Do I need to describe (in a sentence) what dynamical tunneling is?  Or should I let the article take care of that.  I'm afraid by explaining it I may just put extraneous information in}.

%{map: yes, you should briefly state what dynamic tunnelling is; otherwise, readers will just get confused; also, I don't understand how this is related to chaos-assisted tunnelling; when we talk for a few minutes, you should bring Tomsovic's paper and explain this to me }

We have observed a related phenomenon in the rate at which the doubly degenerate eigenvalues at $\alpha=0$ split (see Fig.~\ref{split}).  For example, when $\alpha$ lies strictly between 0 and the first sharp avoided crossing in Fig.~\ref{split}, the $17$th and $18$th levels are regular (with Husimi structure similar to panel A of Fig.~\ref{h_ac}), whereas the $19$th and $20$th levels are chaotic (with Husimi structure similar to panel D of Fig.~\ref{h_ac}).  The eigenvalues of the two chaotic states clearly diverge from each other faster than those of the two regular states.  This observation is reminiscent of that discussed above for avoided crossings except that the eigenenergies are degenerate at some value of $\alpha$ rather than nearly degenerate.  Indeed, suppose that one is starting from the point of closest approach of an avoided crossing.  As $\alpha$ increases, a broad avoided crossing between two states initially localized primarily in integrable regions will display a much smaller rate of divergence from near-degeneracy than a sharp avoided crossing between two chaotic states.  We have also observed that degenerate eigenstates always split into either two chaotic states or two regular ones and that degenerate chaotic states always seem to split at a faster rate than degenerate regular ones.  
%{TM: I included a sentence making this relation a bit more explicit, I hope this clarifies things.}

\subsection{Signatures of Chaos}

Because ergodicity and exponential divergence of phase-space trajectories can be used to characterize classical chaotic systems, it has been suggested that delocalization in the Husimi distributions of a quantum system is a possible signature of chaos in its classical counterpart \cite{sugita1}.  This has been quantified and studied in numerical investigations \cite{timberlakepaper,sugita2} and is also germane to the system investigated here, as the fraction of phase space with chaotic dynamics in the classical system (\ref{hamnod}) increases with $L$ \cite{bievre}.  Consequently, the delocalization and mixing of Husimi structures, which both become more prominent as the prevalence of broad avoided crossings increases with $L$, seem to be signatures of the chaos in the corresponding classical system.

\section{Conclusions}

In this paper, we investigated  a model system with mixed regular and chaotic dynamics that consists of the quantization of a one-dimensional free particle on a ring coupled to a one-dimensional harmonic oscillator.  By examining the eigenenergies as a function of the system parameters (the coupling strength  $\alpha$ and the relative sizes of the interaction and non-interaction regions) and computing Husimi distributions, we studied the quantum signatures of the mixed dynamics.  We identified key integrable and chaotic structures of Husimi distributions by comparing them with the corresponding classical surfaces of section.  For example, we examined sharp avoided crossings between states localized in chaotic regions and those localized in integrable ones and demonstrated numerically the concomitant complete exchange of their Husimi structure \cite{neumann,timberlakepaper,holder}.  We also showed that an avoided crossing between two mixed states tends to be broader than that between a predominately regular state and a predominately chaotic one.  Furthermore, we found that the $\alpha$-step size required to resolve an avoided crossing is correlated with the extent to which the participating states are localized in the chaotic and integrable portions of phase space.  

As the size of the non-interaction region increases, the avoided crossings broaden and their density increases.  This, in turn, leads to an increase in the number of avoided-crossing clusters, in which the participating energy-level curves do not fully exchange slopes (as a function of $\alpha$).  We showed numerically that such avoided-crossing clusters mix the Husimi structures between participating states rather than exchange them fully as in the sharp avoided crossings.  Such mixing tends to promote delocalization in the eigenstates as the coupling strength is increased.  This causes a nontrivial modification in the aggregate Husimi structure as the coupling strength is varied, in contrast to the preservation of the aggregate structure that is characteristic of sharp avoided crossings.  Consequently, as the length of the non-interaction region increases, one observes an increasing amount of mixing in the aggregate Husimi structure as a function of the coupling strength.  This is a signature of the dynamics of the corresponding classical system, for which the chaotic portion of phase space increases with the size of the non-interaction region.  Thus, the appearance of broad avoided crossings, eigenstate delocalization, and the mixing of Husimi structures seem to be signatures of chaos in the quantum system.  Our numerical computations also suggest that the sharpness of avoided crossings is positively correlated with the extent to which the participating Husimi structures are localized in different regions of phase space.  Therefore, the dynamics of avoided crossings in quantum systems seems to be strongly related to the chaotic dynamics of their classical counterparts.

\section*{Acknowledgments}

We thank Stephan De Bi\`evre, Jens Marklof, Jerry Marsden, Alex McCauley, Kevin Mitchell, Paul Parris, and Alex Silvius for useful discussions concerning this research.  We also thank Alex Barnett, Mark Dykman, Joachim Stolze, and two anonymous referees for providing numerous critical comments and suggestions regarding this manuscript.  TM acknowledges support from Caltech's Summer Undergraduate Research Fellowship (SURF) Program and the Richter Memorial Funds.  MAP acknowledges support from the Gordon and Betty Moore Foundation through Caltech's Center for the Physics of Information (where he held a postdoctoral position during this research).

\section*{Appendix I: The Hamiltonian Matrix}

Let ${\cal H}_1={\cal H}_{part}$ be the Hilbert space for a free particle traversing a ring of length $2+L$ and ${\cal H}_2={\cal H}_{osc}$ be the Hilbert space for a harmonic oscillator.  Define $|n \rangle_1  = |\psi^{part}_n \rangle$ and $E^1_n$, respectively, to be the $n$th eigenstate and corresponding $n$th eigenenergy for the particle. We calculate the coordinate-space projections $\{ \psi^{\text{part}}_n(q),\psi^{part}_{-n} \}^{\infty}_{n=1}$ of $\{ |n \rangle_1,|-n \rangle_1 \}^{\infty}_{n=1}$ and their eigenenergies from the Schr\"{o}dinger equation with periodic boundary conditions:
\begin{align*}
	&\frac{\partial^2}{\partial q^2}\psi^{part}_n(q)=-E^{1}_n\psi^{part}_n(q)\,,  \nonumber \\
	&\psi^{part}_n(q+l(2+L))=\psi^{part}_n(q)\,, \: l \in \mathbb{Z}\,.
\end{align*}
The (normalized) solutions are
\begin{align}
	\psi^{part}_n(q)&=\frac{1}{\sqrt{2+L}} \exp\left\{{\left(\frac{2 \pi ni}{2+L}\right)q}\right\}\,, \nonumber \\
E_n^1&=\frac{4 \pi^2 n^2}{\left( 2+L \right)^2}\,, \quad n \in \mathbb{Z} - \{0\} \,.  \label{parteig}
\end{align}
%Here $n \in \mathbb{Z}-\{0\}$ (the $n=0$ state does not exist).  
The ground-state energy of the particle is $E_{\pm 1}^1=4 \pi^2/[( 2+L)^2]$.  
%Note that each eigenspace is two dimensional as the two orthogonal eigenstates $|n \rangle_1$ and $|-n \rangle_1$ correspond to the same eigenenergy.  

Define $| k \rangle_2  = |\psi^{osc}_k \rangle$ and $E_k^2$ to be the $k$th eigenstate and corresponding $k$th eigenenergy of the oscillator.  Here, $E_k^2=k+1/2$, so that $E_0^2=1/2$ is the ground-state energy \cite{cohen-tannoudji}.  
%(The coordinate projection for the eigenstates will not be given here, as it is easier to work with the abstract state vector $|k \rangle_2$.) 
Using the operator definitions (\ref{op}), we write
\begin{align*}
	a^{\dagger} = \frac{1}{\sqrt{2}} \left( \Phi-i\Pi \right)\,, \qquad a = \frac{1}{\sqrt{2}} \left( \Phi+i\Pi \right),
\end{align*}
which are, respectively, the creation and annihilation operators for the harmonic oscillator \cite{cohen-tannoudji, shankar}.  The Hamiltonian (\ref{hamquant}) becomes
\begin{equation}
	H= \left( a^{\dagger}a +\frac{1}{2} \right) - \frac{p^2}{2} -\frac{\alpha}{\sqrt{2}} \left( a^{\dagger}+a \right) \chi(q). \label{hamop}
\end{equation}
The matrix representation of (\ref{hamop}) in the uncoupled basis ${|n \rangle_1 \otimes |k \rangle_2}$ [with integer indices $n \in (-\infty,-1] \cup [1,\infty)$ and $k\in[0,\infty)$]
%^{\infty, \infty}_{n=1,k=0}$ 
is
\begin{equation}
	{\bf H}={\bf E_1} \otimes {\mathbb I} + {\mathbb I} \otimes {\bf E_2} - \alpha {\bf W_1} \otimes {\bf W_2}\,, \label{hammat}
\end{equation}
where ${\mathbb I}$ is the identity matrix and
\begin{align*}
	\left( {\bf E_1} \right)_{nn'}&=\left\langle n\left|\frac{-p^2}{2}\right|n' \right\rangle_1\,, \notag \\
	\left( {\bf E_2} \right)_{kk'}&=\left\langle k\left|\left(a^{\dagger}a +\frac{1}{2}\right)\right|k' \right\rangle_2 \,, \\
	\left( {\bf W_1} \right)_{nn'}&=\left\langle n\left|\chi(q)\right|n' \right\rangle_1\,, \notag \\
	\left( {\bf W_2} \right)_{kk'}&=\left\langle k\left|\frac{1}{\sqrt{2}}\left(a^{\dagger}+a \right)\right|k' \right\rangle_2\,.
\end{align*}

By the definition of the uncoupled basis,
\begin{align}
	\left( {\bf E_1} \right)_{nn'} =\frac{4 \pi^2 n^2}{\left( 2+L \right)^2}\delta_{nn'}\,, \quad
	\left( {\bf E_2} \right)_{kk'} =\left(k+\frac{1}{2}\right)\delta_{kk'}\,.
\end{align}
The coordinate-space projections for the free-particle eigenstates yield
\begin{align*}
	\left( {\bf W_1} \right)_{nn'}&=\int^{2+L}_0 {\psi^{part}_n(q)^{*}}\psi^{part}_{n'}(q) \chi(q)\,dq \notag \\
	&=\int^{2}_0 {\psi^{part}_n(q)^{*}}\psi^{part}_{n'}(q)\,dq\,.
\end{align*}
Hence, with Eq.~(\ref{parteig}), we obtain
\begin{equation}
	\left( {\bf W_1} \right)_{nn'}=
	\begin{cases}
		\frac{1}{2 \pi \left(n-n' \right)} \left(-i+ie^{\frac{4 \pi i(n-n')}{2+L}} \right) & \text{if $n \neq n'$}\,,\\
		\frac{2}{2+L} & \text{if $n=n'$}\,.
	\end{cases}
\end{equation}
Finally, the creation/annihilation operator identities
\begin{align}
	a^{\dagger}|k \rangle_2 = \sqrt{k+1}|k+1 \rangle_2\,, \quad a|k \rangle_2&=\sqrt{k}|k-1 \rangle_2
	\label{ident}
\end{align}
give
\begin{equation}
	\left( {\bf W_2} \right)_{kk'}=\frac{1}{\sqrt{2}} \left(\sqrt{k'+1}\delta_{k,k'+1}+\sqrt{k'}\delta_{k,k'-1} \right).
\end{equation}

\section*{Appendix II: The Husimi Distribution}

The Husimi distribution $H_{\psi}(\bar{p},\bar{q},\bar{\Phi},\bar{\Pi})$ of a state $| \psi \rangle$ of the two-dimensional quantum-mechanical system (\ref{hamquant}) is
\begin{equation}
	H_{\psi}(\bar{p},\bar{q},\bar{\Phi},\bar{\Pi})=N|\langle\psi_{(\bar{p},\bar{q},\bar{\Phi},\bar{\Pi}|)}|\psi\rangle|^2\,, \label{Husimidef}
\end{equation}
where $|\psi_{(\bar{p},\bar{q},\bar{\Phi},\bar{\Pi})}\rangle$ is a coherent state localized around $(\bar{p},\bar{q},\bar{\Phi},\bar{\Pi})$ and $N$ is a normalization constant.  We construct a coherent state for the coupled system as the product
\begin{equation}
	|\psi_{(\bar{p},\bar{q},\bar{\Phi},\bar{\Pi})} \rangle=|\psi_{(\bar{p},\bar{q})} \rangle_1 \otimes \ |\psi_	{(\bar{\Phi},\bar{\Pi})} \rangle_2\,,
\end{equation}
where $|\psi_{(\bar{p},\bar{q})} \rangle_1$ is the coherent state for the free particle and $|\psi_{(\bar{\Phi},\bar{\Pi})} \rangle_2$ is the coherent state for the uncoupled harmonic oscillator.  The latter is given by \cite{cohen-tannoudji}
\begin{equation}
	|\psi_{(\bar{\Phi},\bar{\Pi})} \rangle_2=e^{-\frac{1}{2}\left(\bar{\Phi}^2+\bar{\Pi}^2\right)}\sum_{k=0}^	{\infty}{\frac{\left( \bar{\Phi}+i\bar{\Pi} \right)^k}{\sqrt{k!}}|k\rangle_2}\,.
\label{osccoh}
\end{equation}
Because 
%$q \in (2+L){\mathbb S^1}$ [i.e., 
the uncoupled particle system is $(2+L)$-periodic in $q$ and $p\in {\mathbb R}$, the phase space is cylindrical.  Using the procedure of Ref.~\cite{spina} to define $|\psi_{(\bar{p},\bar{q},\bar{\Phi},\bar{\Pi})} \rangle$ for this topology, we require that the coherent state $|\psi_{(\bar{p},\bar{q})} \rangle_1$ satisfies
\begin{equation}
	\langle q|\psi_{(\bar{p},\bar{q})} \rangle_1 = \langle q+l(2+L)|\psi_{(\bar{p},\bar{q})} \rangle_1,\quad l \in 	\mathbb{Z}\,. \label{ccondition}
\end{equation}
One can construct the coherent states $|\psi_{(\bar{p},\bar{q})} \rangle_1$ from Euclidean coherent states $|\tilde{\psi}_{(\bar{p},\bar{q})} \rangle_1$ (which are Gaussian wavefunctions) by wrapping them around the cylinder and summing overlapping portions.  This yields
\begin{equation}
	\langle q|\psi_{(\bar{p},\bar{q})} \rangle_1=C^{\frac{1}{2}}\sum_{l=-\infty}^{\infty}{\langle q + l(2+L)|	\tilde{\psi}_{(\bar{p},\bar{q})} \rangle_1}\,, \label{coproj}
\end{equation}
which satisfies (\ref{ccondition}) and converges because $\langle q + l(2+L)|\tilde{\psi}_{(p,q)} \rangle_1$ is Gaussian.  In Eq.~(\ref{coproj}), the quantity $C$ is a normalization constant to be determined by the condition $\langle \psi_{(\bar{p},\bar{q})}|\psi_{(\bar{p},\bar{q})}\rangle_1=1$.

The projection of $|\psi_{(\bar{p},\bar{q})} \rangle_1$ onto the uncoupled particle basis $\{|n \rangle_1,|-n \rangle_1 \}_{n=1}^{\infty}$ is
\begin{align*}
	&\langle n|\psi_{(\bar{p},\bar{q})} \rangle_1=\int_0^{2+L}{\langle n|q \rangle_1 \langle q|\psi_{(\bar	{p},\bar{q})} \rangle_1 \, dq} \notag \\
	&=C^{\frac{1}{2}}\sum_{l=-\infty}^{\infty}\int_0^{2+L} {\langle n|q+l(2+L) \rangle_1}   
	 \langle q+l(2+L)|\tilde{\psi}_{(\bar{p},\bar{q})} \rangle_1\, dq\,.
\end{align*}
Using Eq.~(\ref{parteig}) and the $(2+L)$-periodicity of $\psi^{part}_n(q)=\langle q|n \rangle_1$, we obtain
\begin{equation}
	\langle n|\psi_{(\bar{p},\bar{q})} \rangle_1=\int_{-\infty}^{\infty}\langle n|q \rangle_1 \langle q|\tilde{\psi}_{(\bar{p},\bar{q})} \rangle_1 \,dq\,. \label{reduc}
\end{equation}
The coordinate-space projection of the Euclidean coherent state is
\begin{equation}
	\langle q|\tilde{\psi}_{(\bar{p},\bar{q})} \rangle_1=\left(\frac{1}{\pi}\right)^{\frac{1}{4}}\exp\left(-\frac{1}{2}	\left(q-\bar{q}\right)^2+i\bar{p}
	\left(q-\frac{\bar{q}}{2}\right)\right)\,. \label{ecoproj}
\end{equation}
Using the expression for $\langle q|n \rangle_1$ from (\ref{parteig}), we evaluate (\ref{reduc}) and obtain
%Because $\langle n|q+k(2+L) \rangle_1=\psi^{part}_n(q+k(2+L))$, equations (\ref{parteig}), (\ref{coproj}), and (\ref{ecoproj}) give
\begin{equation}
	\langle n|\psi_{(\bar{p},\bar{q})}\rangle_1=C^{\frac{1}{2}}\left(\frac{1}{2}\right)^{{\frac{1}{4}}}\exp\left	({-\frac{1}{2}\left(n-\bar{p}\right)^2-i\bar{q}\left(n-\frac{\bar{p}}{2}\right)}\right)\,. \label{partproj}
\end{equation}
With the normalization condition $\langle \psi_{(\bar{p},\bar{q})}|\psi_{(\bar{p},\bar{q})}\rangle_1=1$, we determine from (\ref{partproj}) that
%with $n \in \mathbb{Z}-\{0\}$ that
\begin{equation}
%	C(\bar{p}) = \sqrt{\pi}\sum_{n \neq 0}e^{-\left(n-\bar{p}\right)^2} \,,
	C = \sqrt{\pi}\sum_{n \neq 0}e^{-\left(n-\bar{p}\right)^2} \,.
\label{norm}
\end{equation}
%where we have emphasized the dependence of $C$ on $\bar{p}$ by writing $C = C(\bar{p})$.  

Thus, if a state $|\psi \rangle$ is expressed in the uncoupled basis as $|\psi \rangle=\sum_{n \neq 0}\sum_{k=0}^{\infty}{a_{nk}|n \rangle_1 \otimes |k \rangle_2}$, we obtain from (\ref{Husimidef}), (\ref{osccoh}), and (\ref{partproj}) that
\begin{align}
	&H_{\psi}(\bar{p},\bar{q},\bar{\Phi},\bar{\Pi})=\frac{C(\bar{p})}{\sqrt{2}}\left | \sum_{n \neq 0}\sum_{k=0}^{\infty}{a_{nk}^{*}	{\frac{\left( \bar{\Phi}+i\bar{\Pi} \right)^k}{\sqrt{k!}}}} \right. \notag \\
	 &\times \left. \exp\left({-\frac{1}{2}\left[\bar{\Phi^2}+\bar{\Pi}^2+\left(n-\bar{p}\right)^2\right]-i\bar{q}\left(n-	\frac{\bar{p}}{2}\right)}\right)\right |^2\,. \label{Husimi}
\end{align}
In practice, we truncate the sum in Eq.~(\ref{Husimi}) in order to compute the Husimi distribution for eigenstates calculated using a truncated Hamiltonian matrix for (\ref{hamquant}).  To compare Husimi distributions with classical Poincar\'e surface of sections, we take $\bar{p}=(2E-\bar{\Phi}^2-\bar{\Pi}^2+2\alpha \bar{\Phi})^{1/2}$ (the eigenstate has energy $E$) and an appropriate value of $\bar{q}$ (with, for example, $\bar{q}=1$ for a section in the center of the interaction region).  The value for $\bar{p}$ arises as a slice along the classical energy shell, and the value for $\bar{q}$ corresponds to the choice of Poincar\'e SOS for the classical system \cite{bievre}.

%\bibliographystyle{plain}
%\bibliography{main}

\end{document}